\documentclass[a4paper]{JHEP3} % 10pt is ignored!

\preprint{\arXivid{1004.2432}}
\usepackage{epsfig,multicol,bbm}
\usepackage[abs]{overpic}
\newcommand\fverb{\setbox\fverbbox=\hbox\bgroup\verb}
\newcommand\fverbdo{\egroup\medskip\noindent%
			\fbox{\unhbox\fverbbox}\ }
\newcommand\fverbit{\egroup\item[\fbox{\unhbox\fverbbox}]}
\newbox\fverbbox

\newcommand{\shellfont}[1]{{\fontfamily{pcr}\selectfont\texttt{#1}}}

\newcommand{\mosesWithSpace}[0]{{\sc{Moses }}}

\newcommand{\pythiaWithSpace}[0]{{\sc {Pythia}}8 }
\newcommand{\pythiaNoSpace}[0]{{\sc {Pythia}}8}
\newcommand{\pythia}[0]{{\sc Pythia}8}
\newcommand{\shat}{\mbox{$\sqrt{\hat{s}}(\mu^+\mu^-)$}}
\newcommand{\slhc}{\mbox{$\sqrt{s}_{\rm LHC}$}}
\newcommand{\llhc}{\mbox{${\cal L}_{\rm LHC}$}}

\title
{A search for heavy Kaluza-Klein electroweak gauge bosons at the LHC}

\author{{Gideon Bella, Erez Etzion, Noam Hod, Yaron Oz, Yiftah Silver}\\
{The Raymond and Beverly Sackler School of Physics and Astronomy, Tel-Aviv University}\\
E-mail: \email{noam.hod@cern.ch}}

\author{{Mark Sutton}\\
{The Department of Physics and Astronomy, University of Sheffield}\\
E-mail: \email{sutt@cern.ch}}

\abstract{
The feasibility for the observation of a 
certain leptonic Kaluza-Klein (KK) hard process in {\em pp} 
interactions at the LHC is presented. 
Within the $S^1/Z_2$ TeV$^{-1}$ extra dimensional 
theoretical framework with the focus on the KK excitations of 
the Standard Model $\gamma$ and $Z^0$ gauge bosons, the 
hard-process, 
$f\bar f \to \sum_n\left(\gamma^*/Z^*\right)_n \to F \bar F$,
has been used where $f$ is the initial state parton,
$F$ the final state lepton and $\left(\gamma^*/Z^*\right)_{n}$ is the $n^{\rm th}$ KK excitation 
of the $\gamma/Z^0$ boson.
For this study the analytic form for the hard process cross section has 
been independently calculated by the authors and has been implemented 
using the {\sc Moses} framework.
The Moses framework itself, that has been written by the authors, was used as an external process 
within the {\pythia}
Monte Carlo generator
which provides the phase space generation for the final state leptons and  
partons from the initial state hadrons, and the simulation of 
initial and final state radiation and hadronization. A brief discussion 
of the possibility for observing and identifying the unique signature of the KK 
signal given the current LHC program is also presented.
}


\newcommand{\figone}{
\FIGURE[tth]{
\epsfig{file=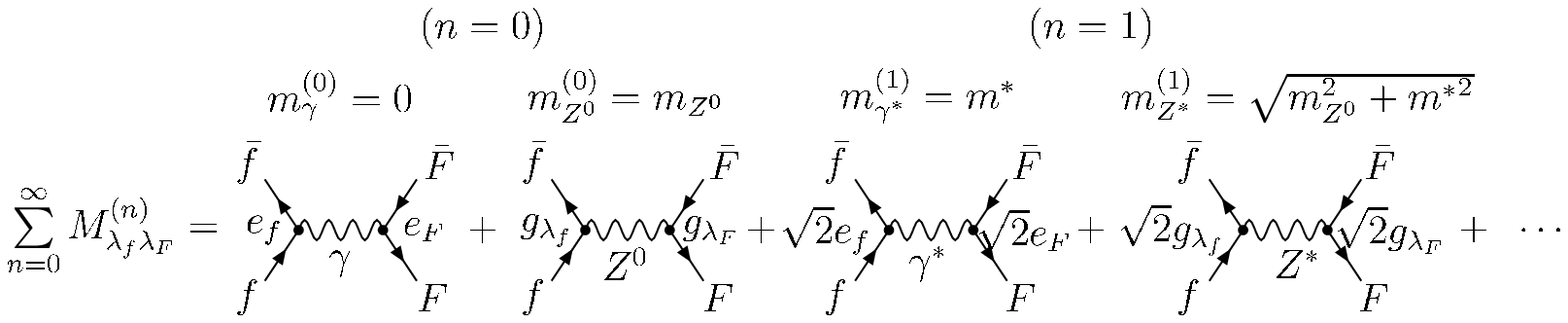,width=15cm}
\vspace{-0.5cm}
\caption{%\textsl
{The Feynman diagram of the entire KK tower of the 
$\gamma/Z^0$ gauge bosons.}}%starting from the $0^{th}$ SM state.}}
\label{fig:KK_feynman}
}
}

\newcommand{\figtwo}{
\FIGURE{
\begin{minipage}{15cm}
\begin{minipage}{7.5cm}
	\begin{overpic}[scale=0.39,angle=0]{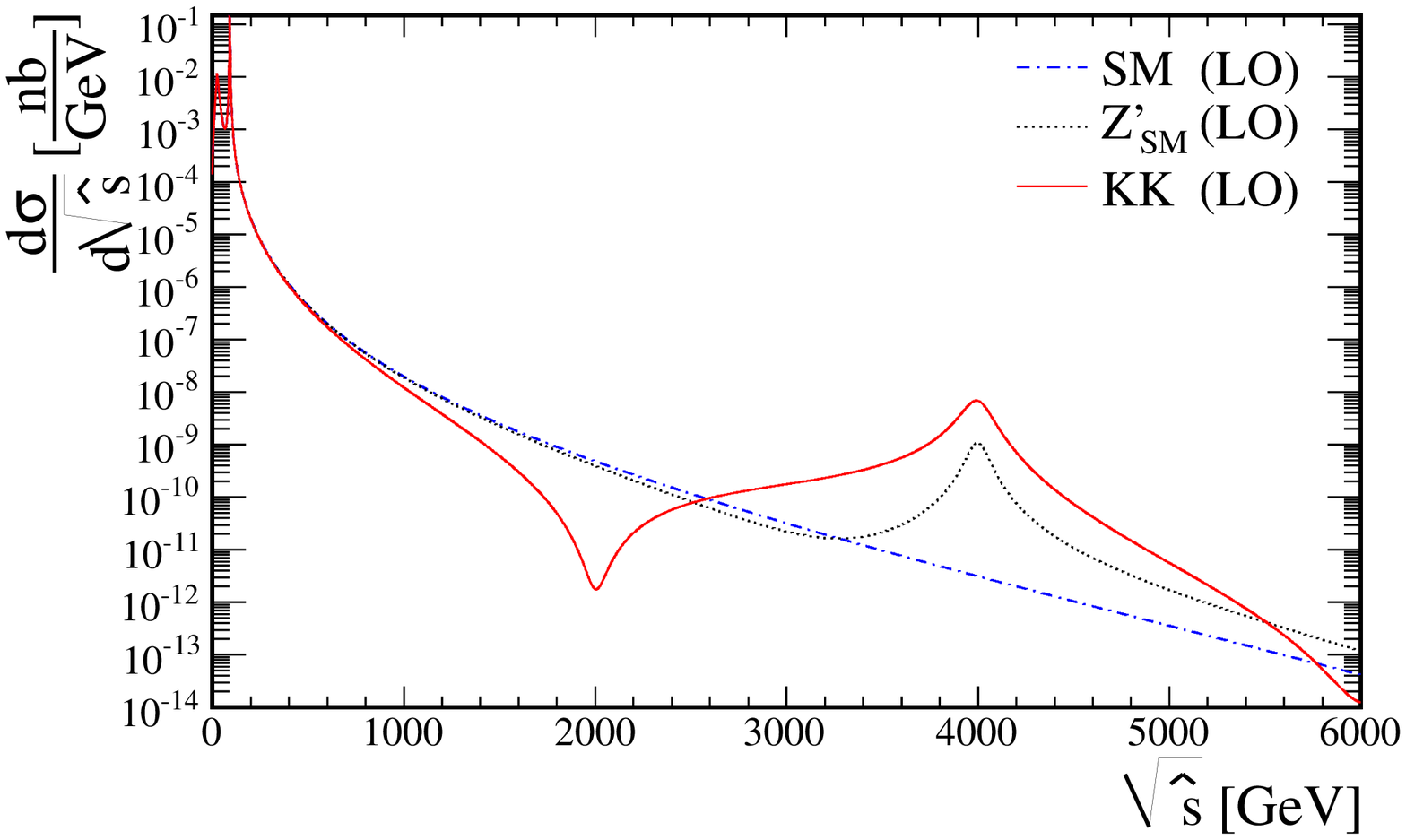}
		\put(60,45){(a)}
	\end{overpic}
\end{minipage}
\begin{minipage}{7.5cm}
	\begin{overpic}[scale=0.39,angle=0]{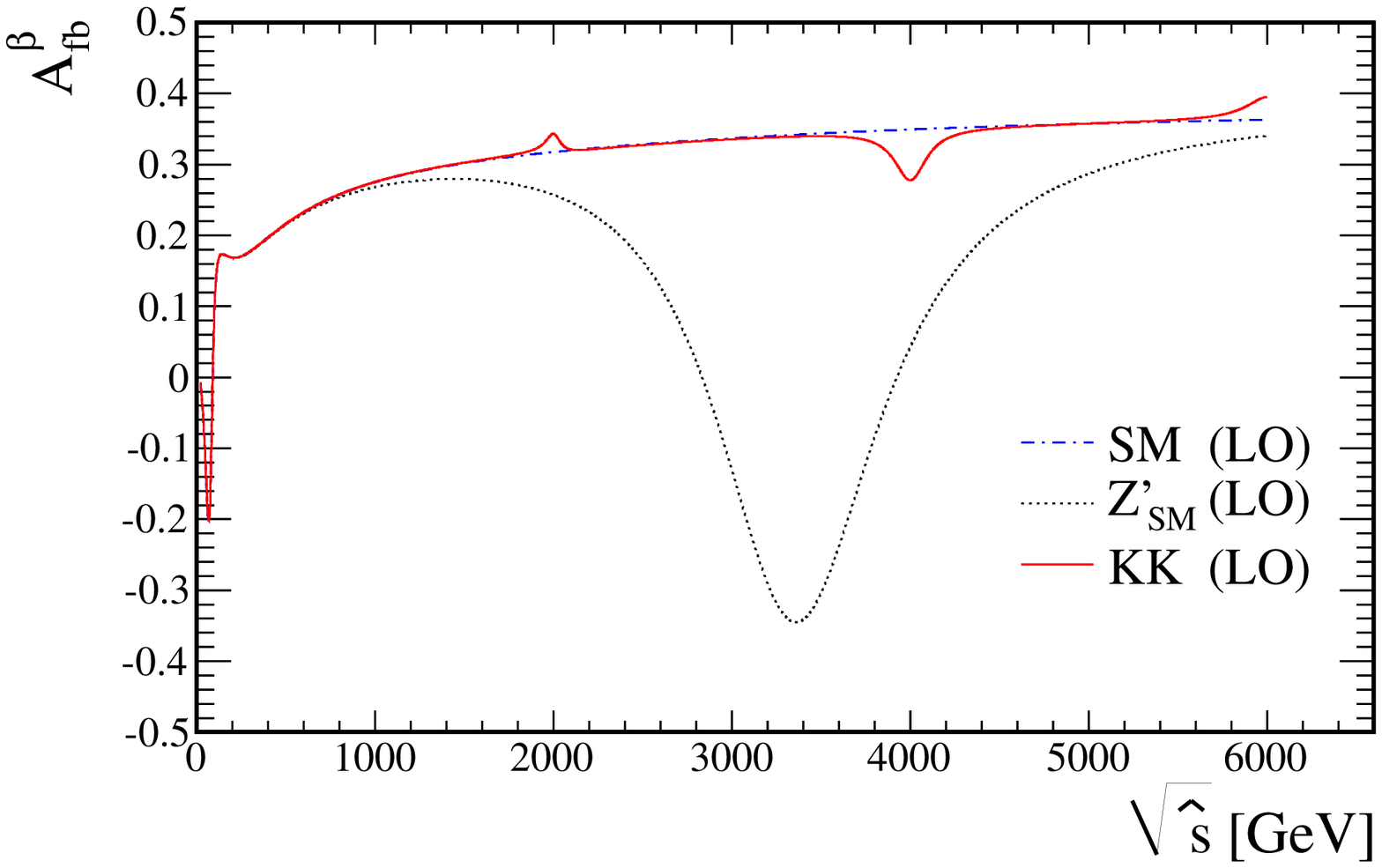}
		\put(60,45){(b)}
	\end{overpic}
\end{minipage}
\end{minipage}
\vspace{-0.1cm}
\caption{%\textsl
{The leading order (LO) invariant-mass distribution {\rm{(a)}} and the Forward-Backward 
asymmetry {\rm{(b)}} for the three models, 
KK (solid), $Z'_{{\rm{SM}}}$ (dotted) and SM (dash-dot) 
discussed in the text.}}
\label{fig:theory_3models}
}
}

\newcommand{\figthree}{
\FIGURE{
\hspace{-0.6cm}\begin{minipage}{15cm}
\begin{minipage}{7.5cm}\begin{overpic}[scale=0.4]{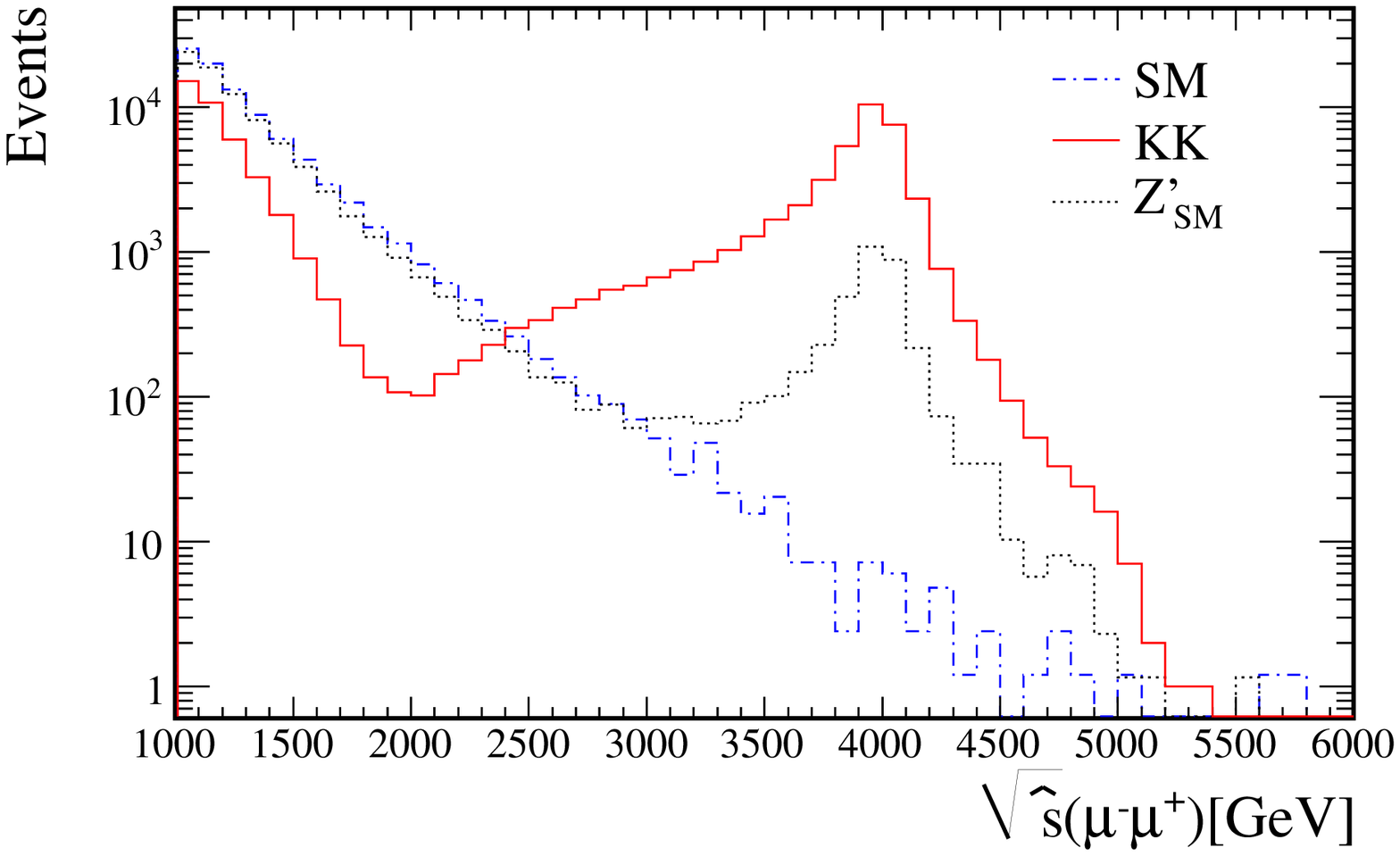}\put(55,40){(a)}\end{overpic}\vspace{2mm}\end{minipage}
\hspace{0.2cm}
\begin{minipage}{7.5cm}\begin{overpic}[scale=0.4]{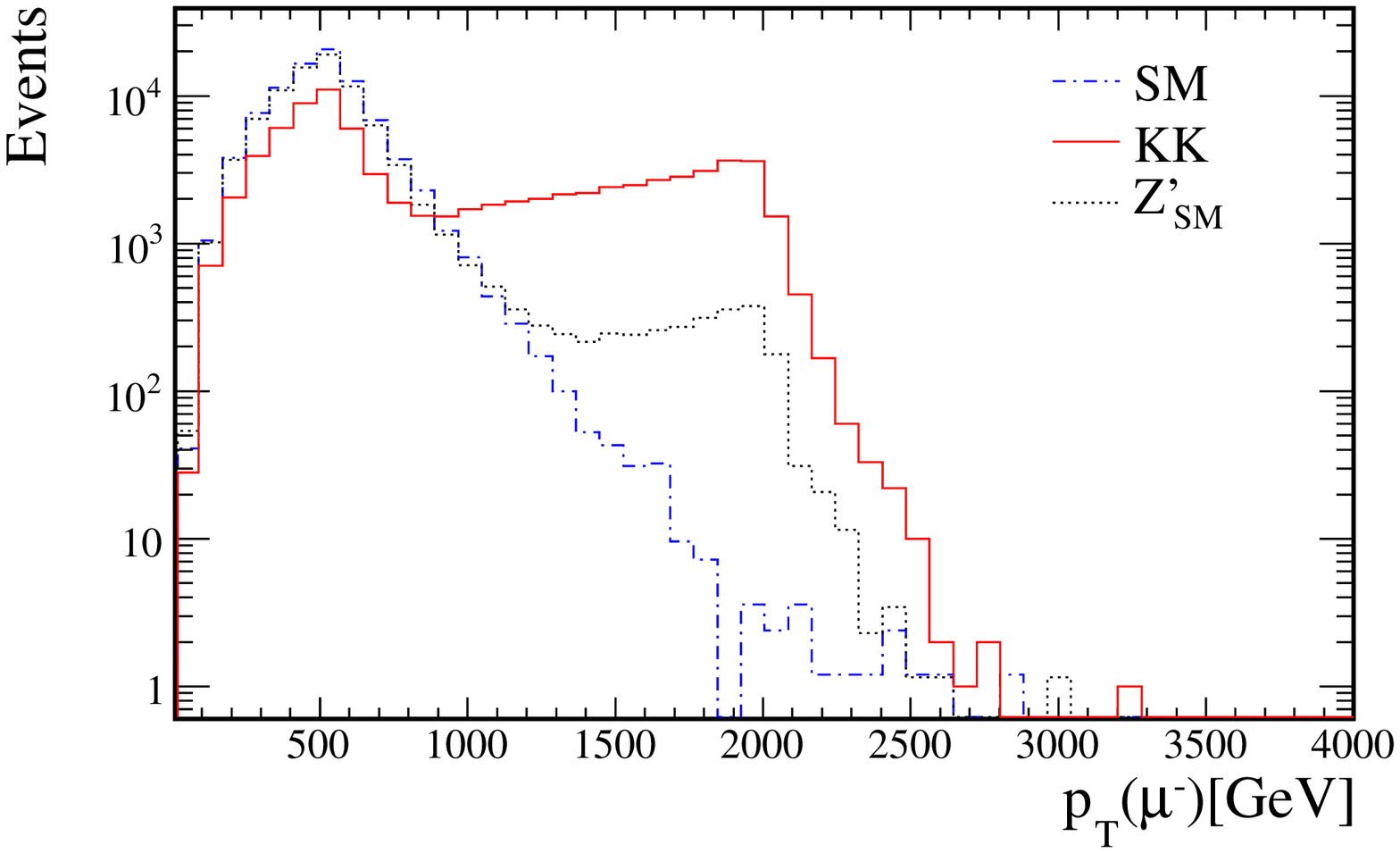}\put(55,40){(b)}\end{overpic}\vspace{2mm}\end{minipage}
\begin{minipage}{7.5cm}\begin{overpic}[scale=0.4]{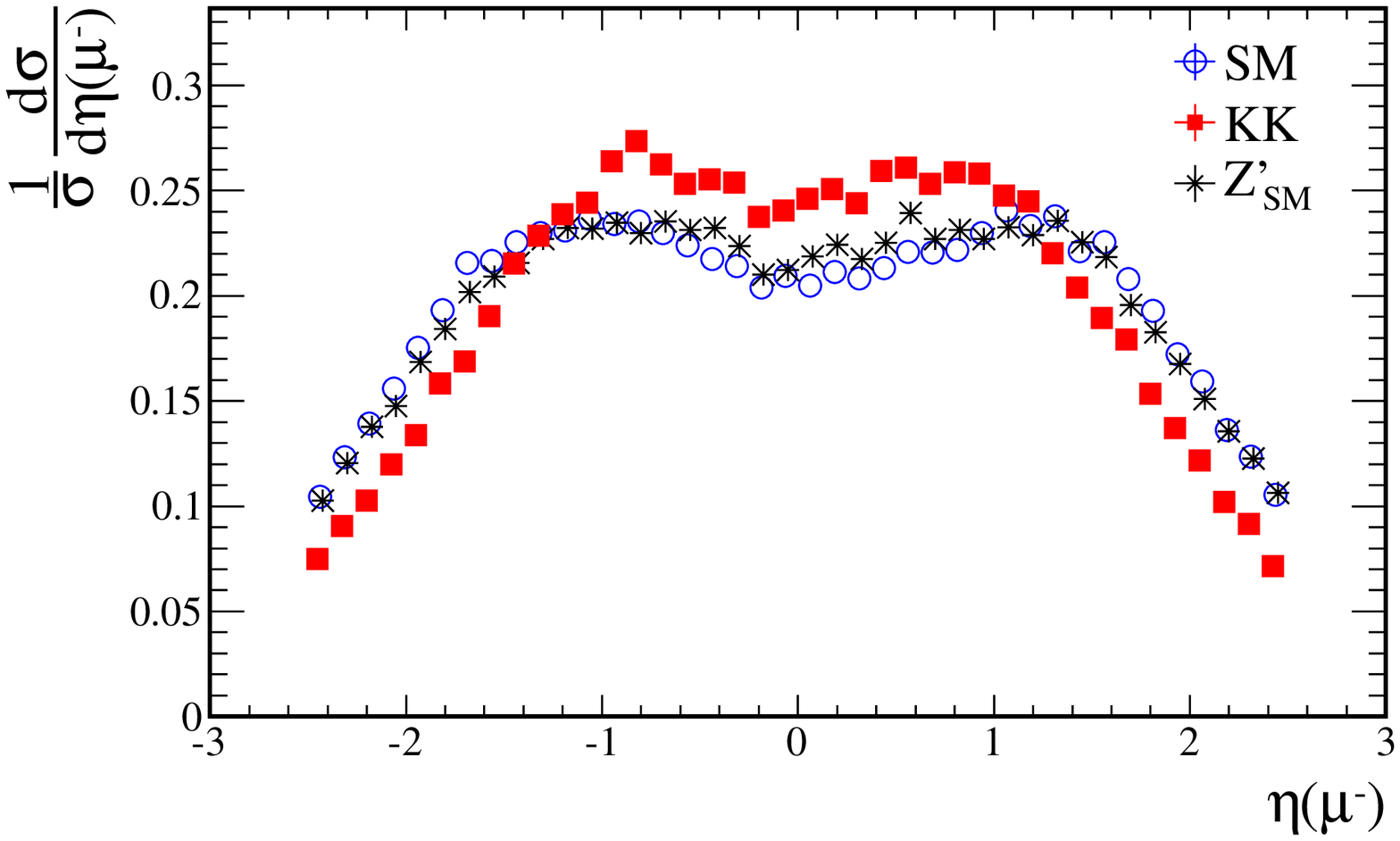}\put(55,40){(c)}\end{overpic}\end{minipage}
\hspace{0.2cm}
\begin{minipage}{7.5cm}\begin{overpic}[scale=0.4]{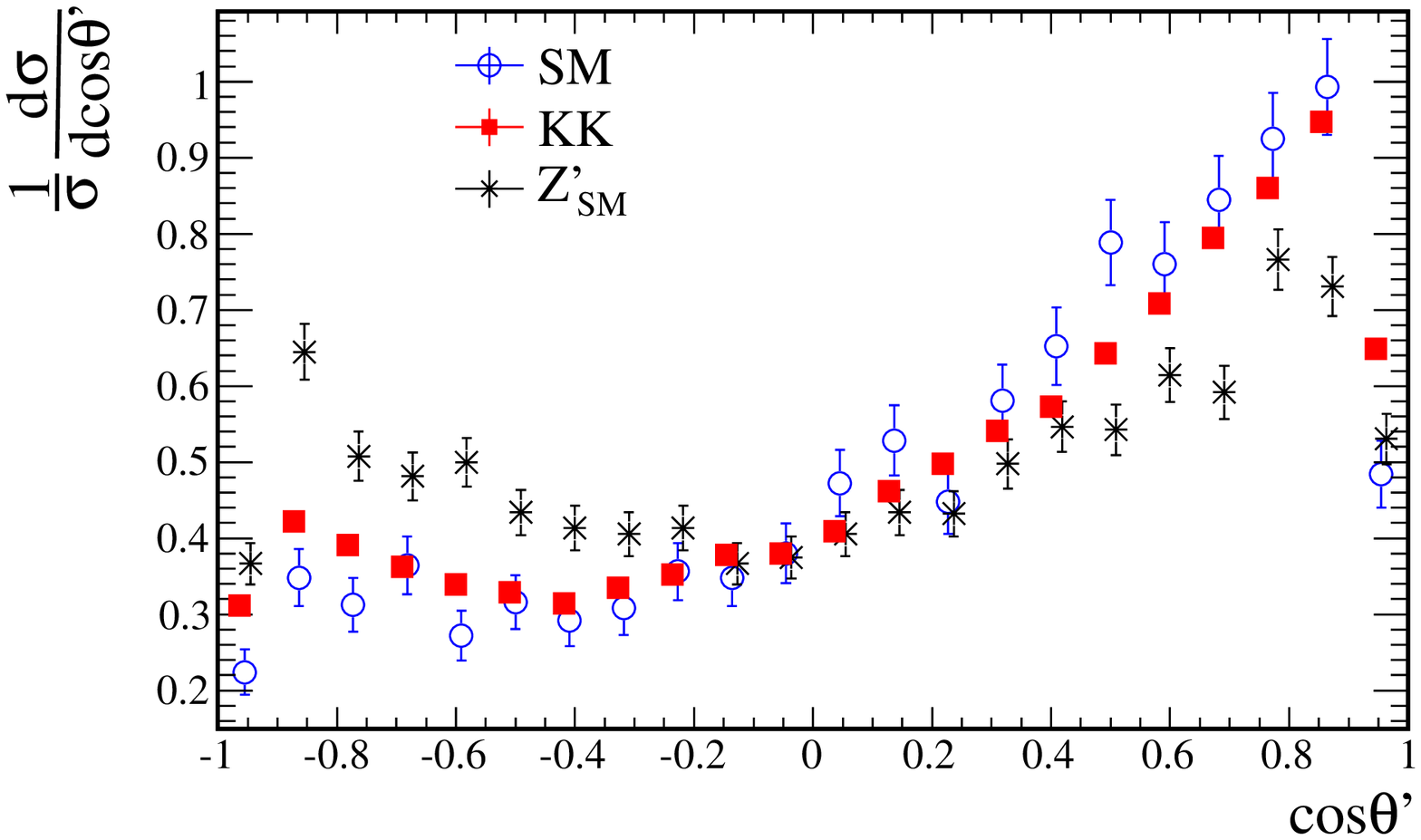}\put(75,85){$2\leq\sqrt{\hat{s}}\leq 5~$TeV}
\put(185,40){(d)}\end{overpic}\end{minipage}
\end{minipage}
\caption{%\textsl
Kinematic distributions of the Monte Carlo reference samples for the KK (solid), 
the SM (dash-dot) and the $Z'_{{\rm{SM}}}$ (dotted) models; 
{\rm (a)} The di-muon invariant mass, \shat, %$\sqrt{\hat s}$; 
{\rm (b)} the muon $p_T$ spectrum,
{\rm (c)} the muon normalised $\eta$ distribution, 
{\rm (d)} the normalised muon 
$\cos\theta'$ distribution. An additional requirement of 
$2\leq\shat\leq 5~$TeV has been applied to the events in 
{\rm (d)} since the forward-backward asymmetry exhibits a 
strong dependence on the invariant mass (see figure~\ref{fig:theory_3models}).}
\label{fig:KK_kinematics_highLuminosity}
}
}

\newcommand{\figfour}{
\FIGURE{
\hspace{-0.6cm}\begin{minipage}{15cm}
\begin{minipage}{7.5cm}\begin{overpic}[scale=0.395]{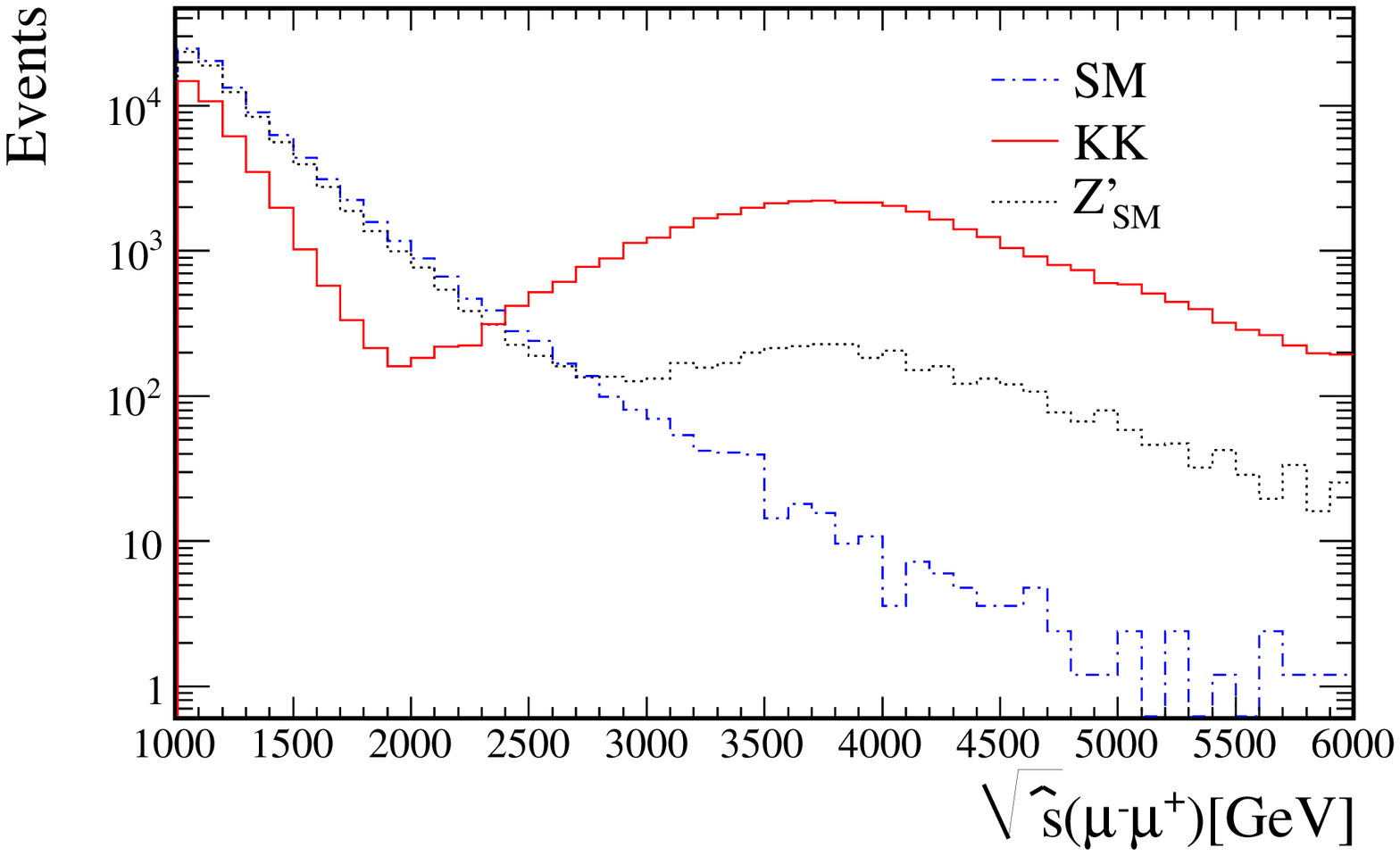}\put(55,40){(a)}\end{overpic}\vspace{2mm}\end{minipage}
\hspace{0.2cm}
\begin{minipage}{7.5cm}\begin{overpic}[scale=0.395]{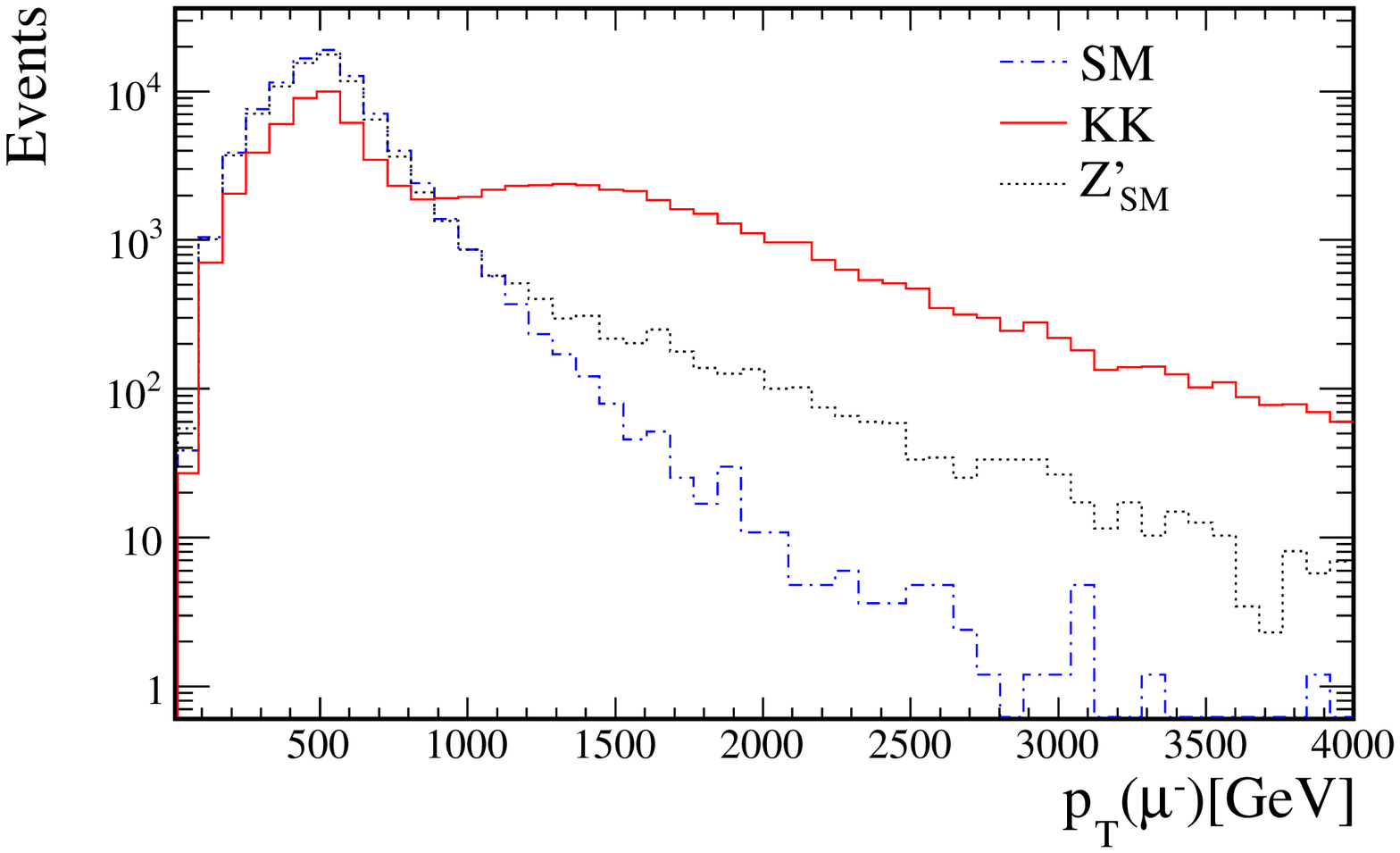}\put(55,40){(b)}\end{overpic}\vspace{2mm}\end{minipage}
\end{minipage}
\vspace{-0.2cm}
\caption{%\textsl
Kinematic distributions of the Monte Carlo reference samples for the KK (solid), 
the SM (dash-dot) and the $Z'_{{\rm{SM}}}$ (dotted) models when including smearing 
to simulate realistic momentum resolution at the detector level; 
{\rm (a)} The reconstructed di-muon invariant mass, \shat %$\sqrt{\hat s}$; 
{\rm (b)} the reconstructed muon $p_T$ spectrum.
}
\label{fig:threeb}
}
}

\newcommand{\figfive}{
\FIGURE{
\hspace{-0.6cm}
\begin{minipage}{15cm}
\begin{minipage}{7.5cm}
	\begin{overpic}[scale=0.39,angle=0]{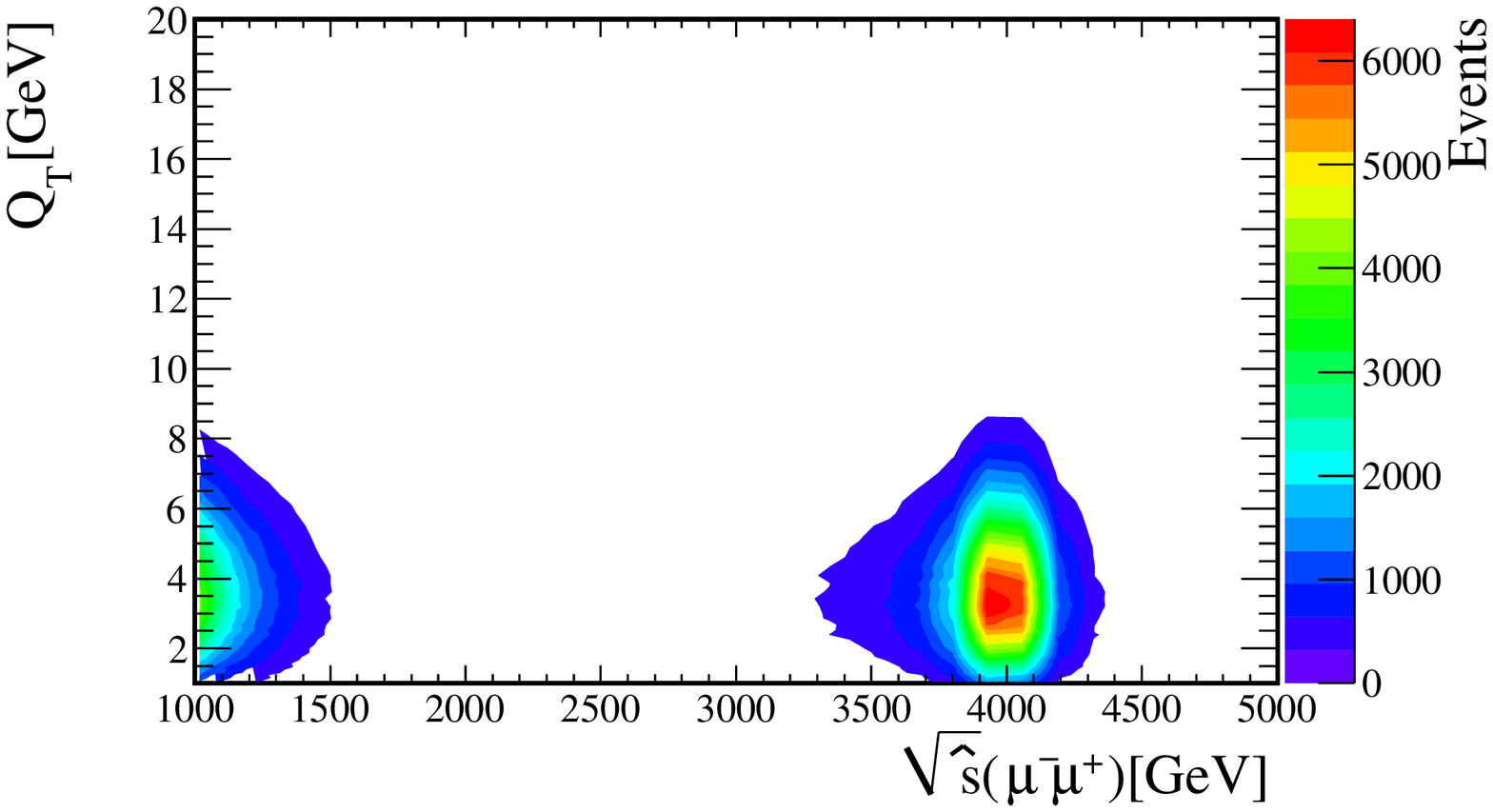}
		\put(45,100){(a)}
	\end{overpic}
\end{minipage}
\hspace{0.2cm}
\begin{minipage}{7.5cm}
	\begin{overpic}[scale=0.39,angle=0]{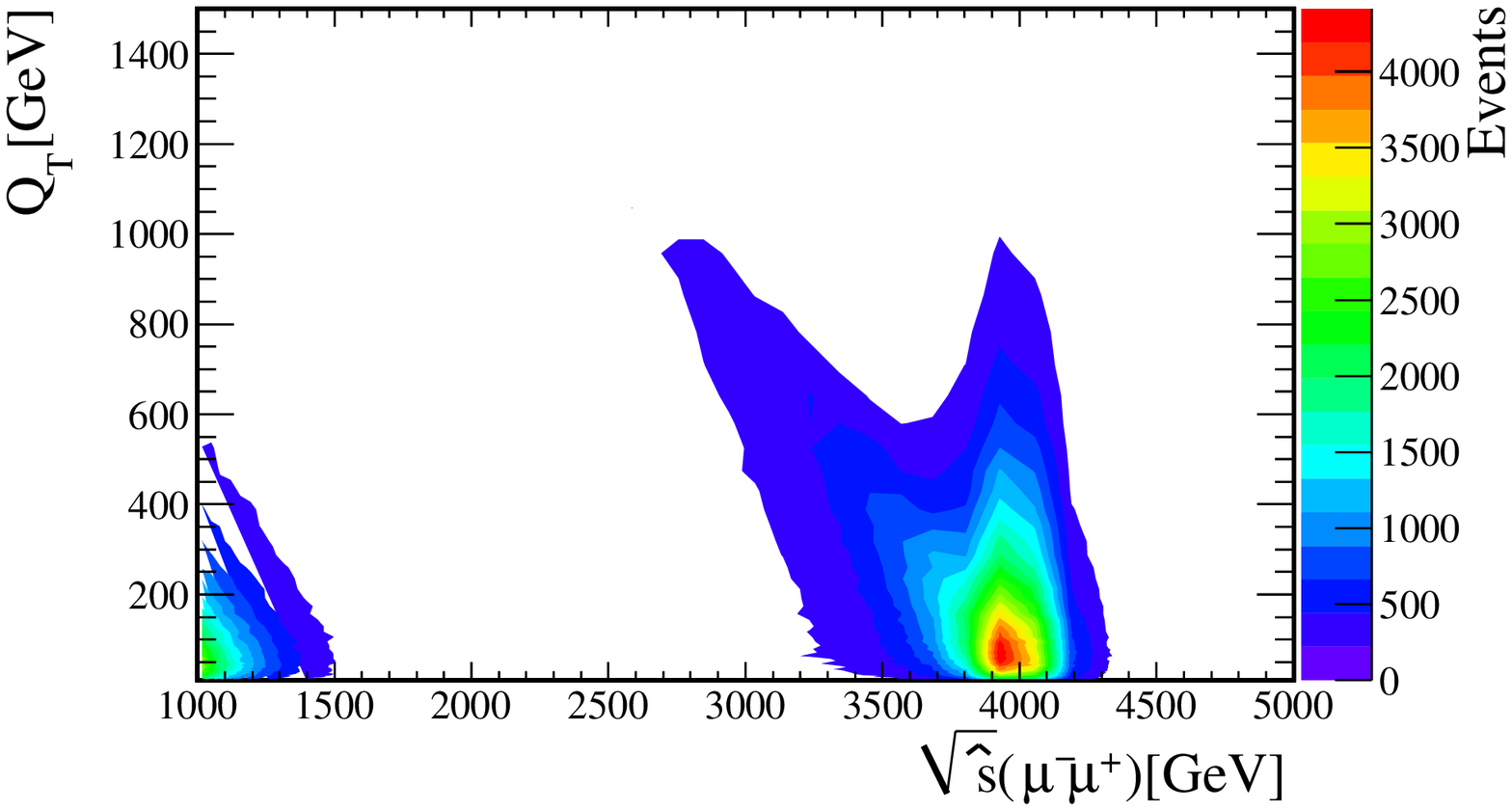}
		\put(45,100){(b)}
		\put(85,87){{\footnotesize $\gamma$ FSR}}
		\put(140,87){{\footnotesize $g$ ISR}}
	\end{overpic}
\end{minipage}
\end{minipage}
\caption{%\textsl
{The distribution of the transverse momentum of the di-muon state, 
$Q_T$ versus the  di-muon invariant mass, $\sqrt{\hat s}$. In {\rm (a)}, the 
initial- and final-state radiation in {\sc Pythia}8 is switched off, 
whereas in {\rm (b)} it is switched on. Note the larger scale for the 
y-axis in {\rm (b)}.}}
\label{fig:QT}
}
}

\newcommand{\figsix}{
\FIGURE{
\hspace{-0.6cm}\begin{minipage}{15cm}
\begin{minipage}{7.5cm}\begin{overpic}[scale=0.4]{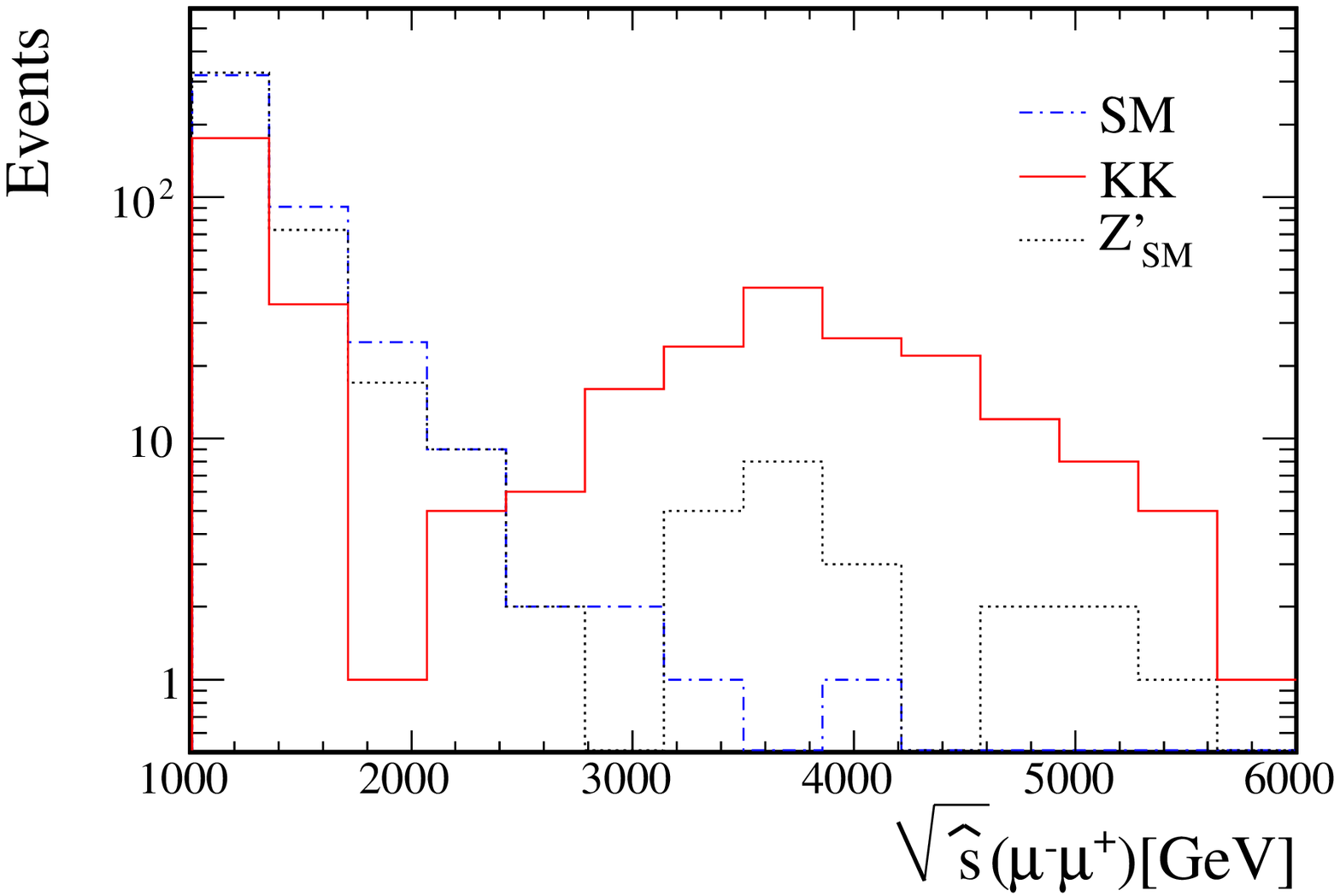}\put(60,122){(a)}\end{overpic}\vspace{2mm}\end{minipage}
\hspace{0.2cm}
\begin{minipage}{7.5cm}\begin{overpic}[scale=0.4]{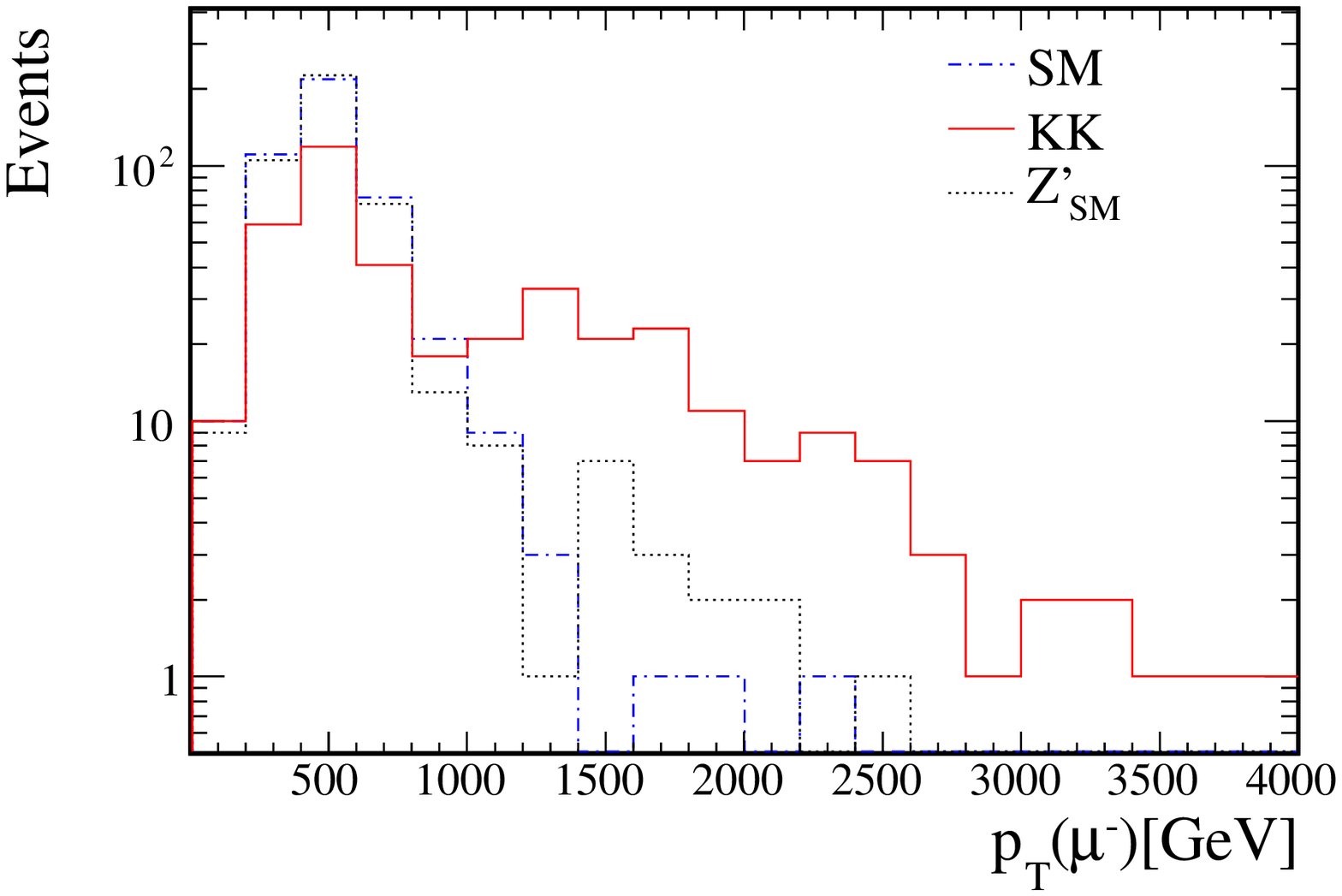}\put(60,40){(b)}\end{overpic}\vspace{2mm}\end{minipage}
\begin{minipage}{7.5cm}\begin{overpic}[scale=0.4]{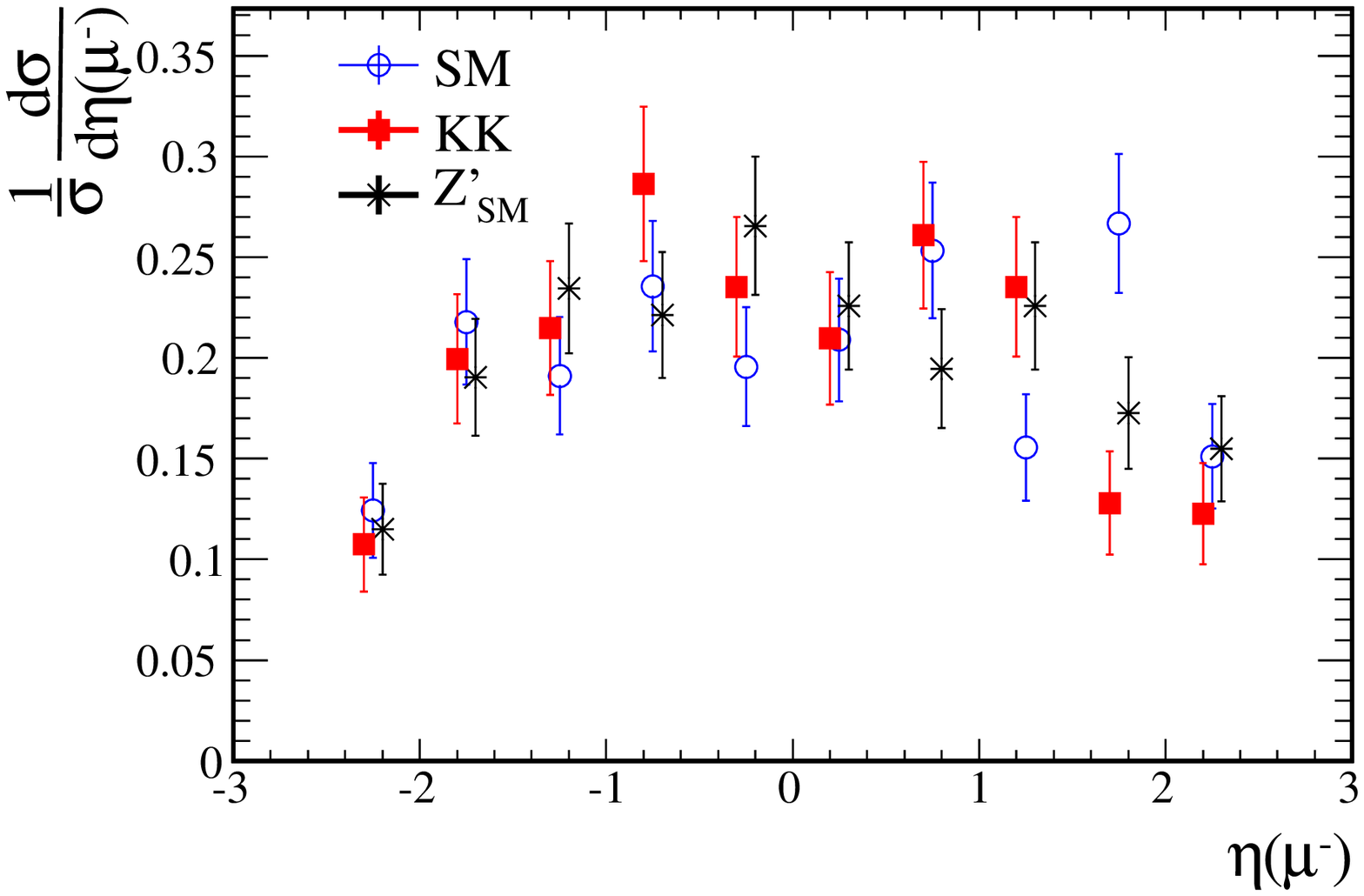}\put(60,35){(c)}\end{overpic}\end{minipage}
\hspace{0.2cm}
\begin{minipage}{7.5cm}\begin{overpic}[scale=0.4]{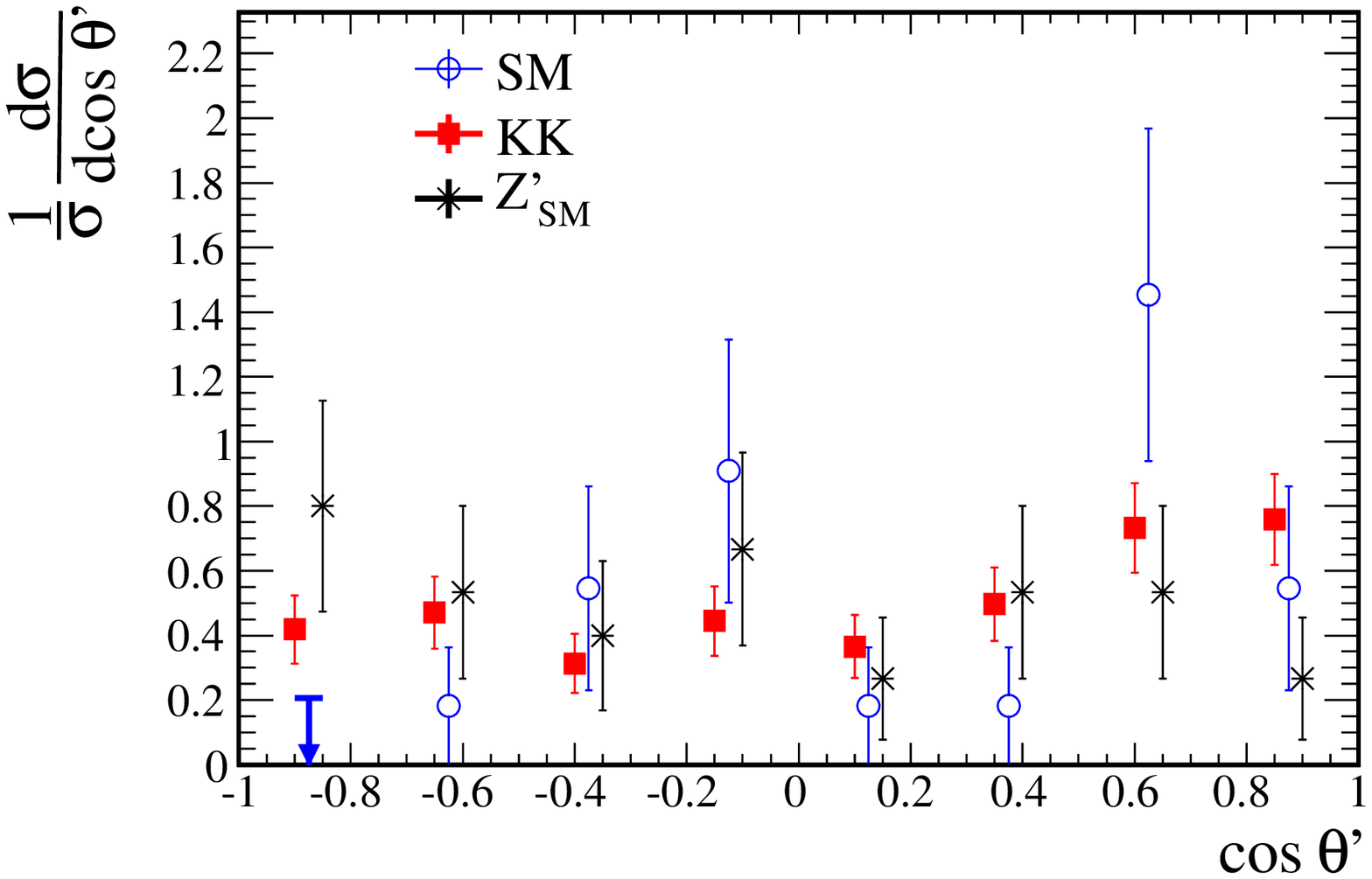}\put(103,122){$2\leq\sqrt{\hat{s}}\leq 5~$TeV}
\put(52,122){(d)}\end{overpic}\end{minipage}
\end{minipage}
\caption{
Kinematic distributions for an integrated luminosity, \llhc=100~fb$^{-1}$ 
for the KK (solid), the SM (dash-dot) and the $Z'_{\rm SM}$ (dotted) models;
{\rm (a)} The di-muon invariant mass, \shat,
{\rm (b)} the muon transverse momentum, $p_T$, 
{\rm (c)} the normalised muon $\eta$ and, 
{\rm (d)} the normalized muons $\cos\theta '$ distributions. 
The additional requirement of 
$2\leq\shat\leq 5$~TeV has been applied to the events in (d)
as in figure~\ref{fig:KK_kinematics_highLuminosity}.
Note in {\rm (d)}, the empty bin for the 
Standard Model distribution at $\cos\theta^\prime$ of -0.9. 
The length of the arrow in this case represents the  
1$\sigma$ Poisson upper limit.}
\label{fig:KK_kinematics_lowLuminosity100}
}
}

\newcommand{\figseven}{ 
\FIGURE{
\hspace{-0.4cm}\begin{minipage}{15cm}
\begin{minipage}{7.5cm}\vspace{0.35cm}\begin{overpic}[scale=0.4]{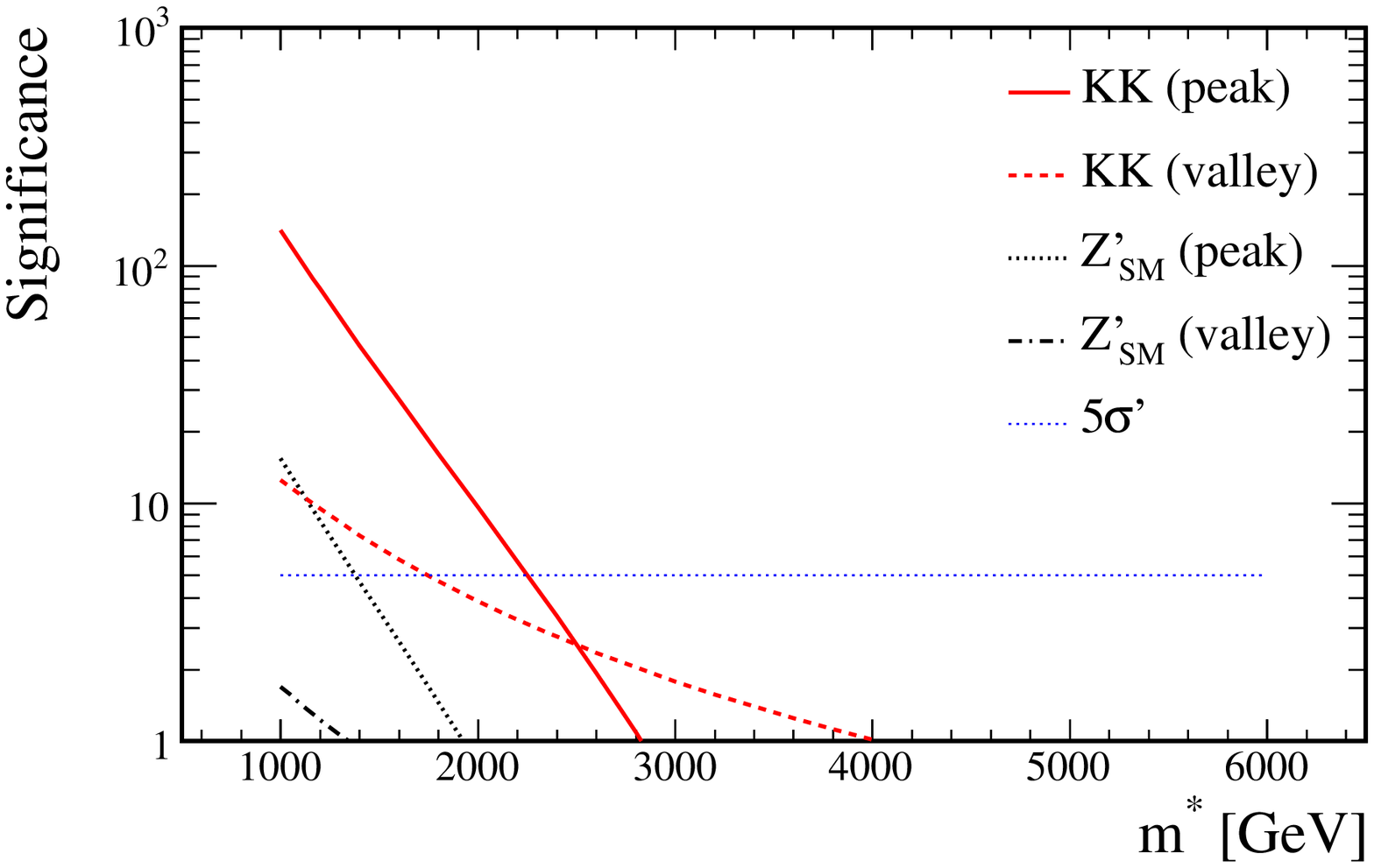}\put(135,115){(a)}
   \put(+40,138){{\footnotesize \slhc=7~TeV, \llhc=1~fb$^{-1}$}}
\end{overpic}\vspace{2mm}\end{minipage}
\hspace{0.1cm}
\begin{minipage}{7.5cm}\vspace{0.35cm}\begin{overpic}[scale=0.4]{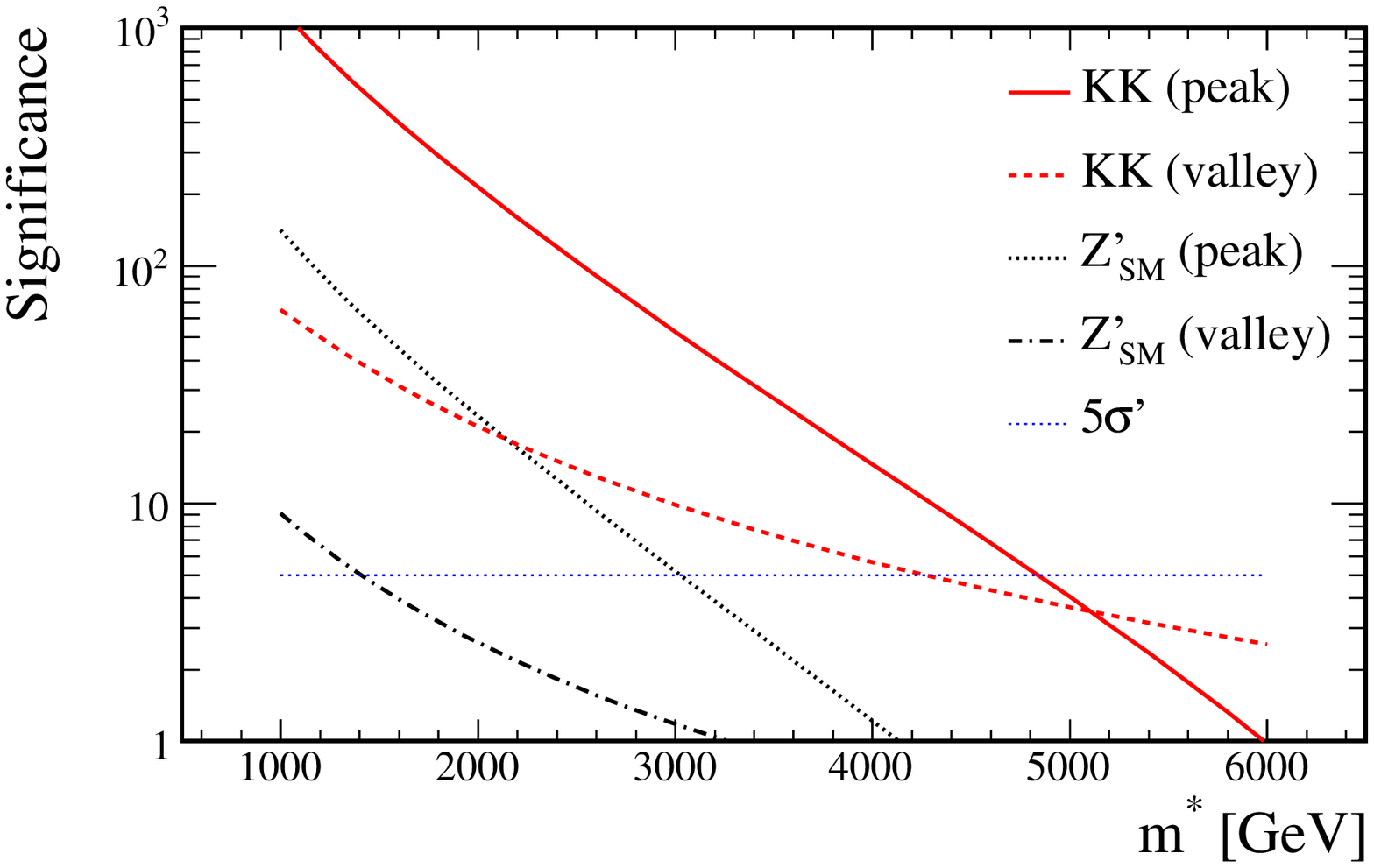}\put(135,115){(b)}
   \put(+40,138){{\footnotesize \slhc=14~TeV, \llhc=10~fb$^{-1}$}}
\end{overpic}\vspace{2mm}\end{minipage}
\begin{minipage}{7.5cm}\vspace{0.35cm}\begin{overpic}[scale=0.4]{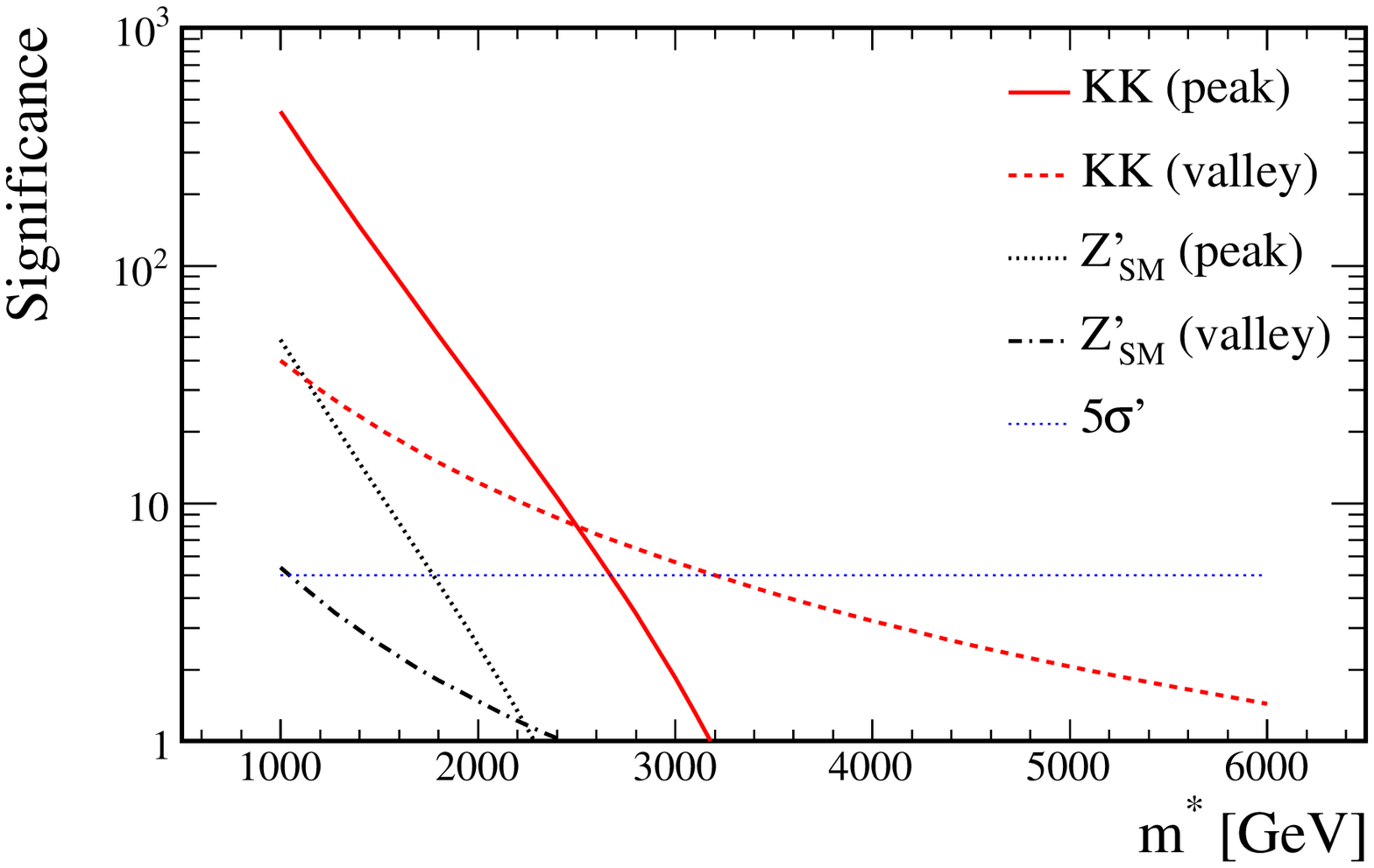}\put(135,115){(c)}
   \put(+40,138){{\footnotesize \slhc=7~TeV, \llhc=10~fb$^{-1}$}}
\end{overpic}\end{minipage}
\hspace{0.1cm}
\begin{minipage}{7.5cm}\vspace{0.35cm}\begin{overpic}[scale=0.4]{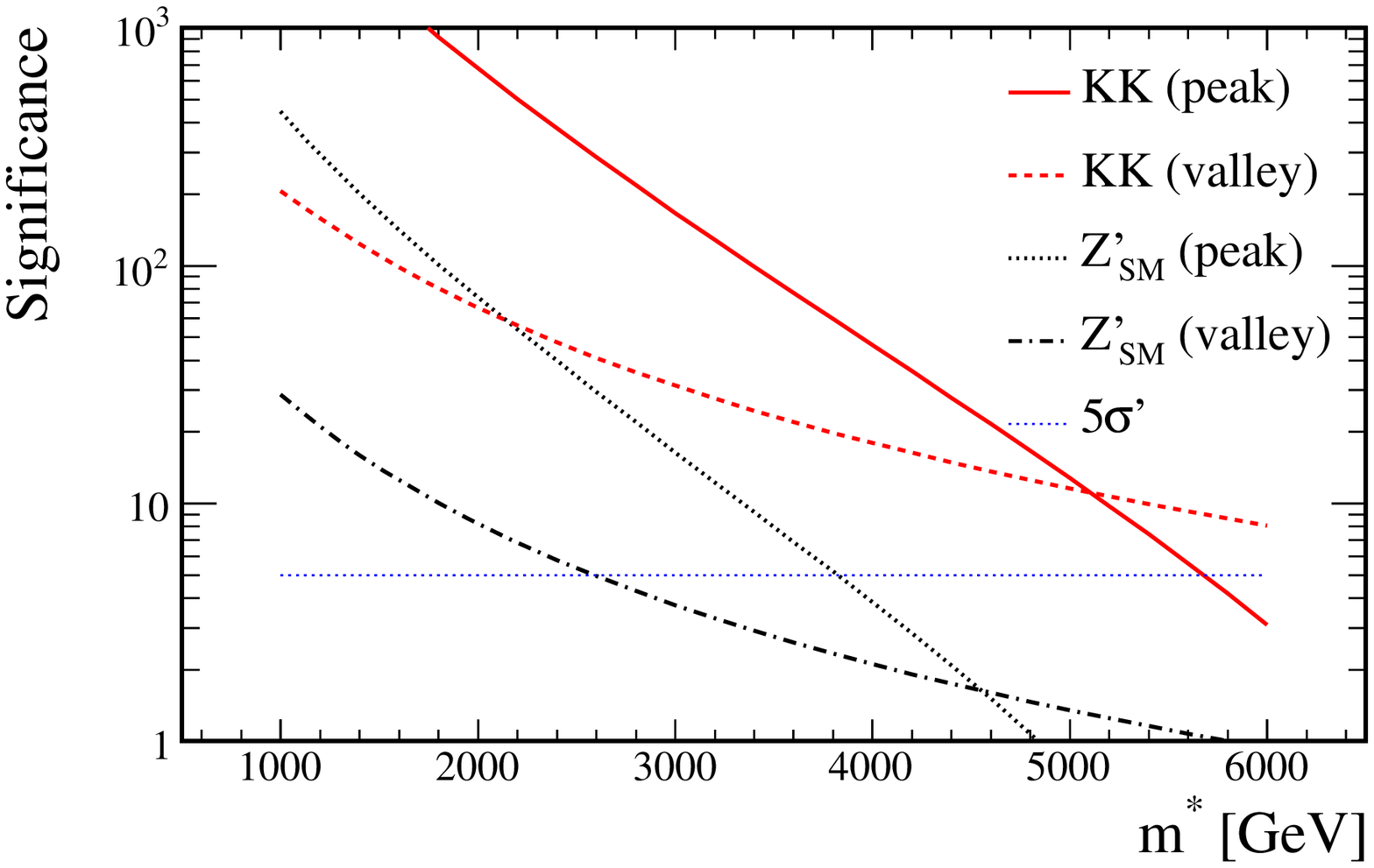}\put(135,115){(d)}    
    \put(+40,138){{\footnotesize \slhc=14~TeV, \llhc=100~fb$^{-1}$}}
\end{overpic}\end{minipage}
\end{minipage}
\caption{
The expected significance for observing events around 
the first KK peak (solid-red) or in the first KK valley region (red-dashed) 
as a function of the BSM mass parameter, calculated using 
equation~\ref{eq:significance} and after an overall K-factor of 1.25 has been applied, the 
effect of which is to improve the mass reach by transposing the significance curves 
approximately 50~GeV towards higher masses with respect to the significance without 
applying this factor.
Also shown is the expected significance for the 
the $Z'_{{\rm{SM}}}$ model calculated in the KK peak (black-dotted) and valley 
(black-dash-dotted) regions, also calculated after applying the 1.25 K-factor. 
For illustration, the horizontal $5\sigma$ line (thin-blue-dotted) is also shown.
}
\label{fig:significance}
}
}

\newcommand{\figeight}{
\FIGURE[tth]{
\begin{minipage}{15cm}
\hspace{-0.8cm}
\begin{minipage}{7.5cm}
	\begin{overpic}[scale=0.4]{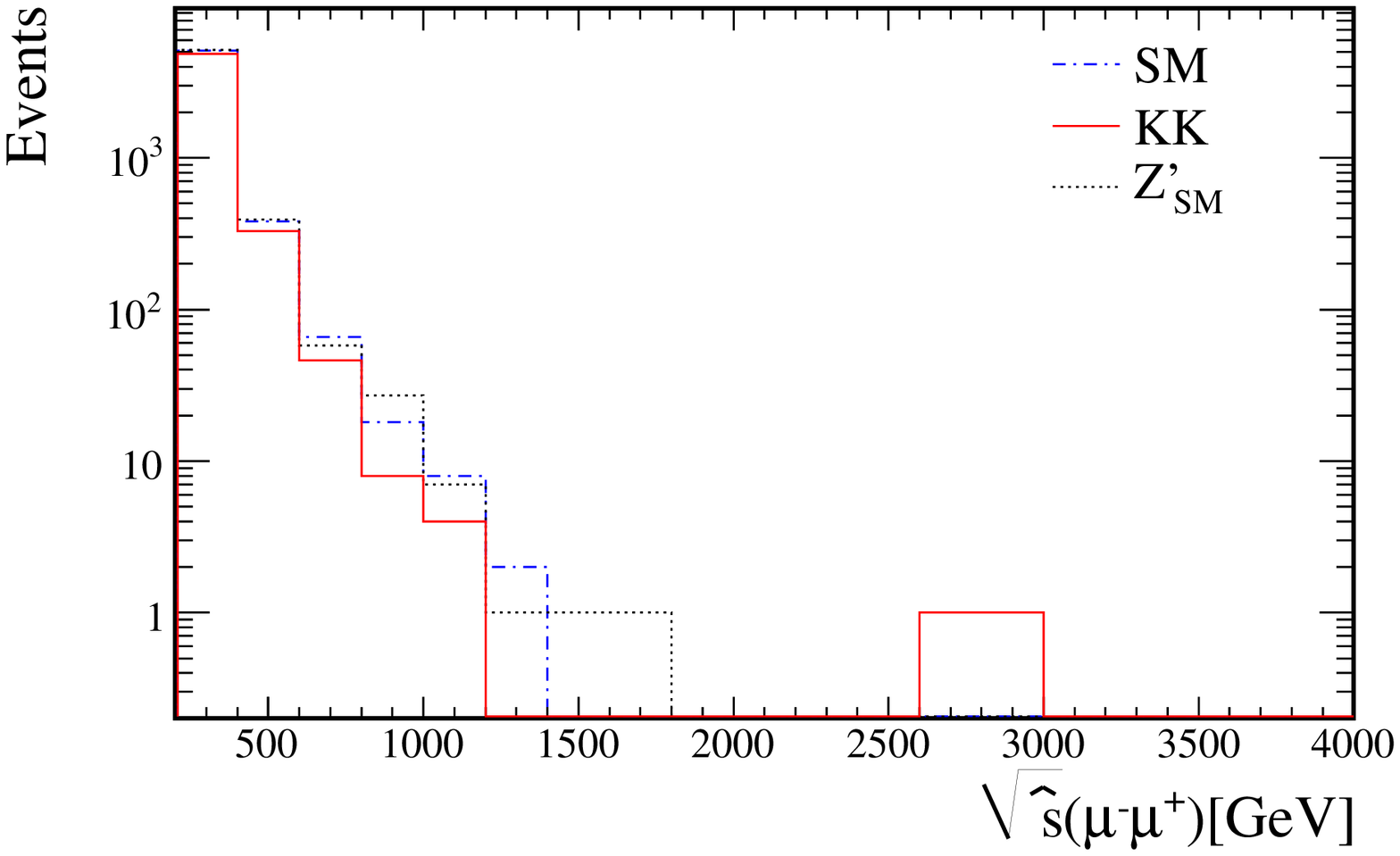}
          \put(140,115){(a)}
          \put(65,115){$m^*=3~{\rm TeV}$}
        \end{overpic}
\end{minipage}
\hspace{0.1cm}
\begin{minipage}{7.5cm}
	\begin{overpic}[scale=0.4]{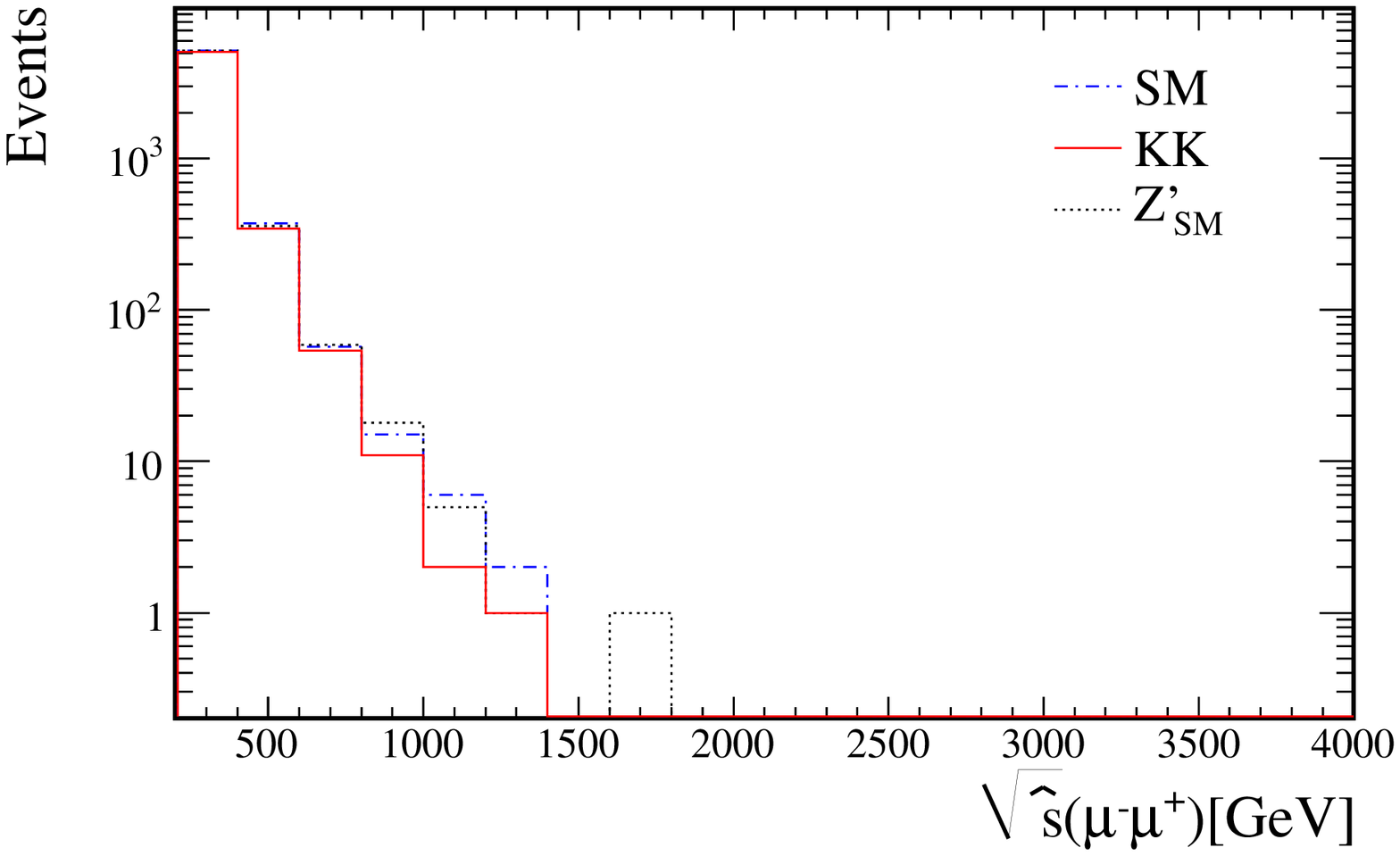}
          \put(140,115){(b)}
          \put(65,115){$m^*=4~{\rm TeV}$}
        \end{overpic}
\end{minipage}
\end{minipage}
\begin{minipage}{15cm}
\hspace{-0.8cm}
\begin{minipage}{7.5cm}
	\begin{overpic}[scale=0.4]{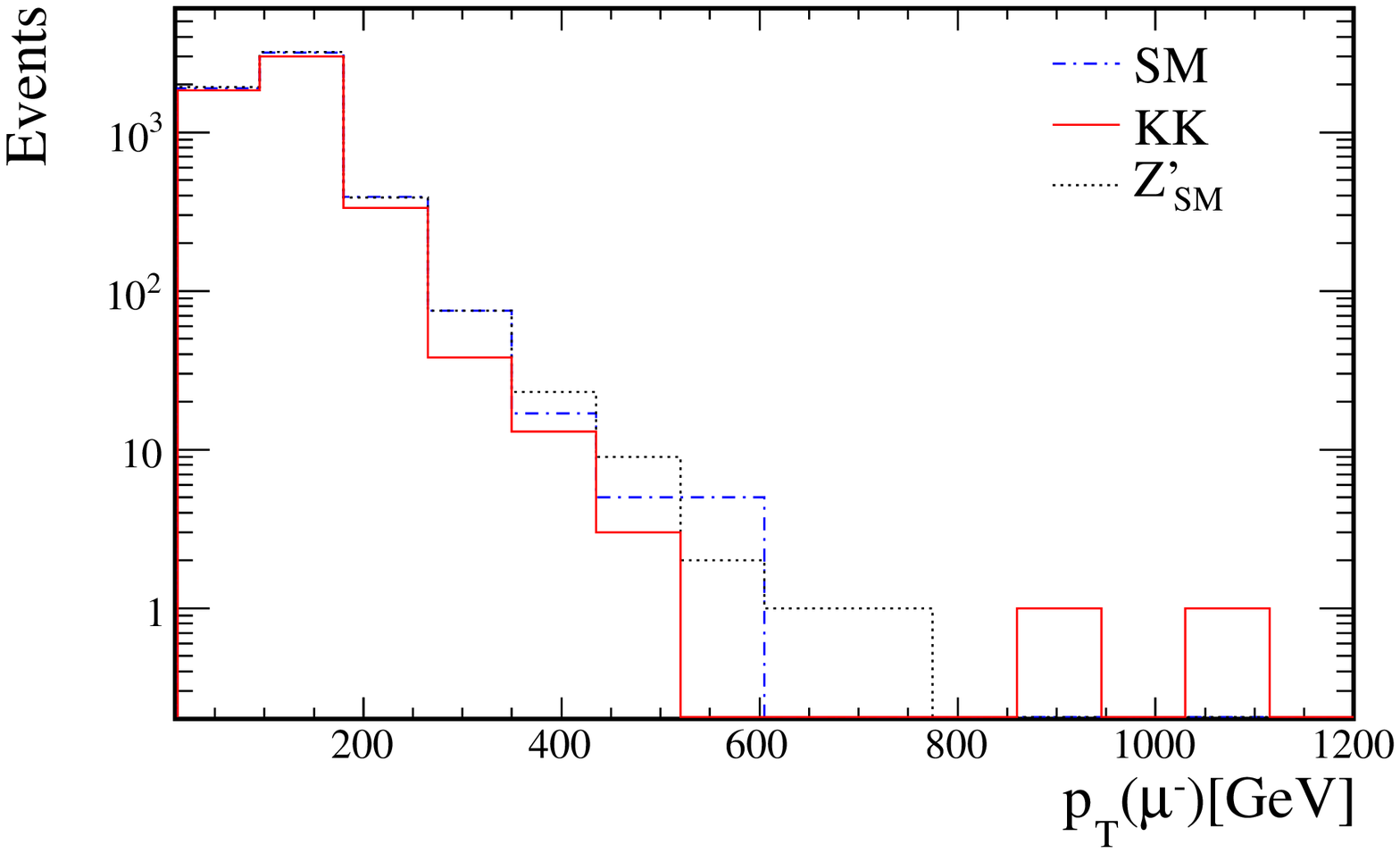}
          \put(140,115){(c)}
          \put(65,115){$m^*=3~{\rm TeV}$}
        \end{overpic}
\end{minipage}
\hspace{0.1cm}
\begin{minipage}{7.5cm}
	\begin{overpic}[scale=0.4]{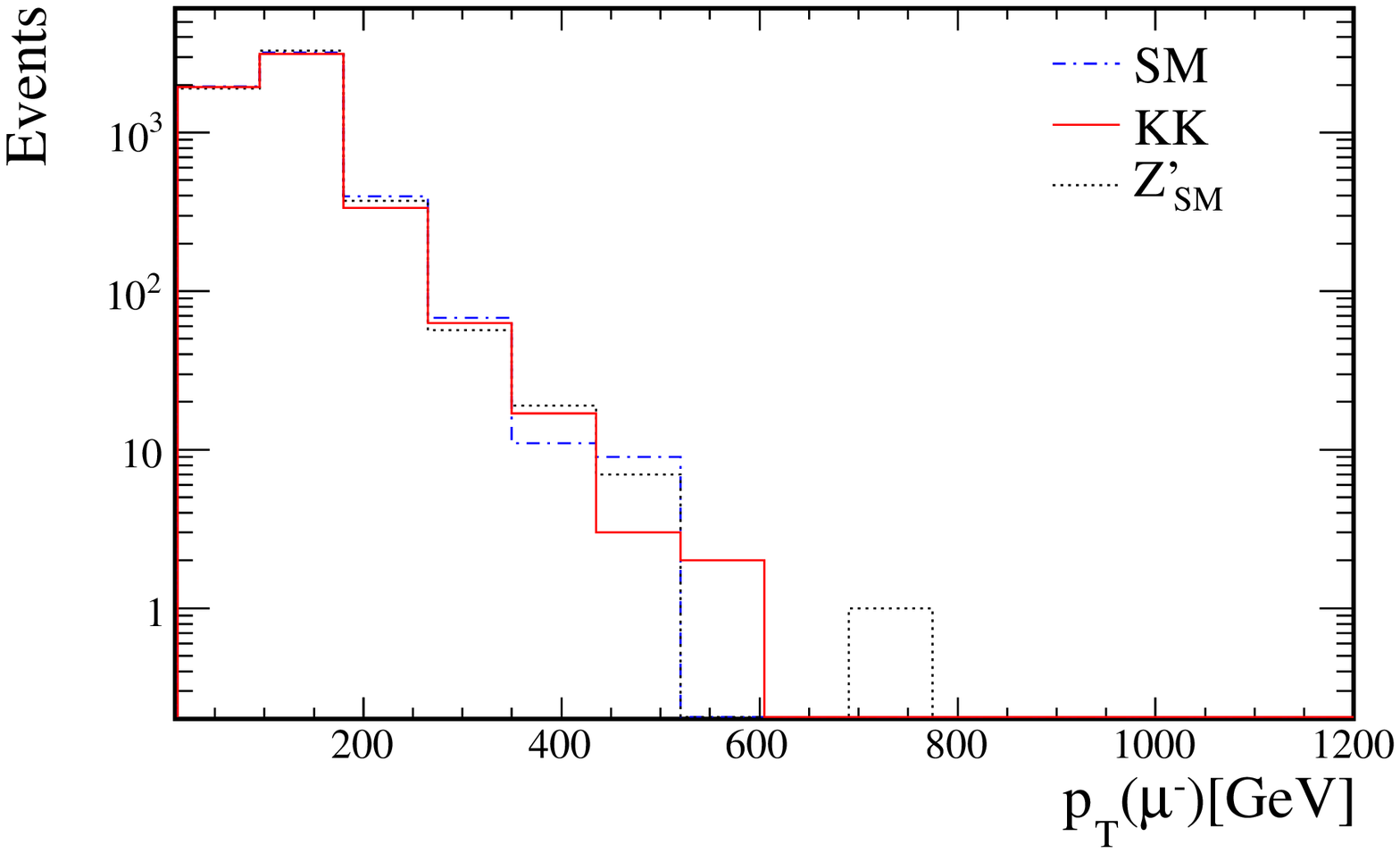}
          \put(140,115){(d)}
          \put(65,115){$m^*=4~{\rm TeV}$}
        \end{overpic}
\end{minipage}
\end{minipage}
\vspace{-0.2cm}
\caption{%\textsl{
Expected distributions corresponding to
\slhc=7~TeV and \llhc=10~fb$^{-1}$ for the KK (red, solid), the Standard Model
(blue, dash-dot) and the $Z'_{\rm SM}$ (black, dotted) models, 
as a function of the di-muon invariant mass, \shat, 
for the two cases of mass parameter, $m^*=3$~TeV (a) 
and $m^*=4$~TeV (b), and as a function of the muon transverse momentum, $p_T(\mu^-)$,
also for $m^*=3$~TeV (c) and $m^*=4$~TeV (d).  
}
\label{fig:binbybin}
}
}

\newcommand{\fignine}{
  \FIGURE[tth]{
    \hspace{-0.6cm}
    \begin{minipage}{15cm}
      \begin{minipage}{7.5cm}
        \begin{overpic}[scale=0.4]{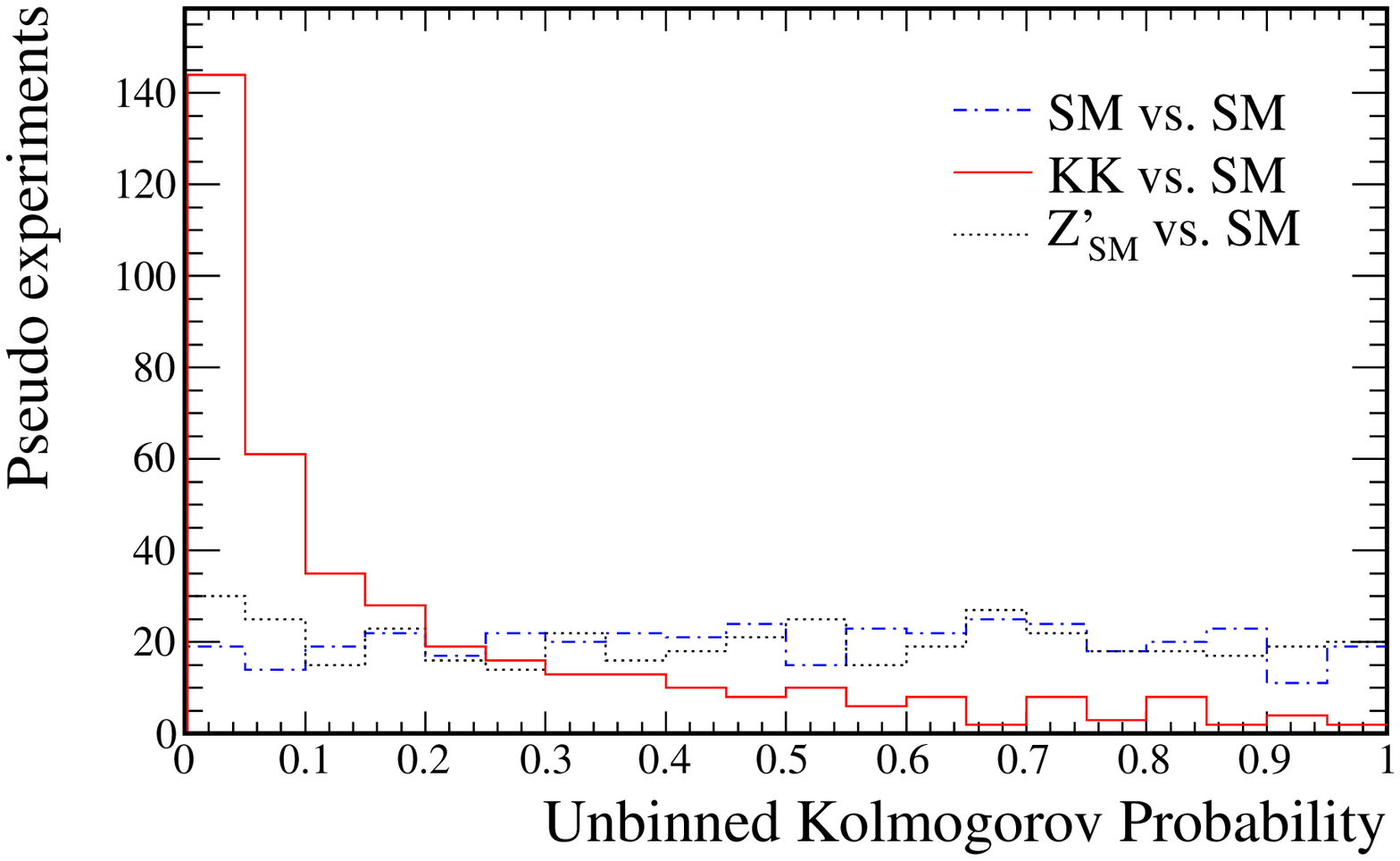}
          \put(60,110){(a) \ $m^*=3$~TeV}\put(60,95){Invariant mass}
        \end{overpic}
      \end{minipage}
      \hspace{1mm}
      \begin{minipage}{7.5cm}
        \begin{overpic}[scale=0.4]{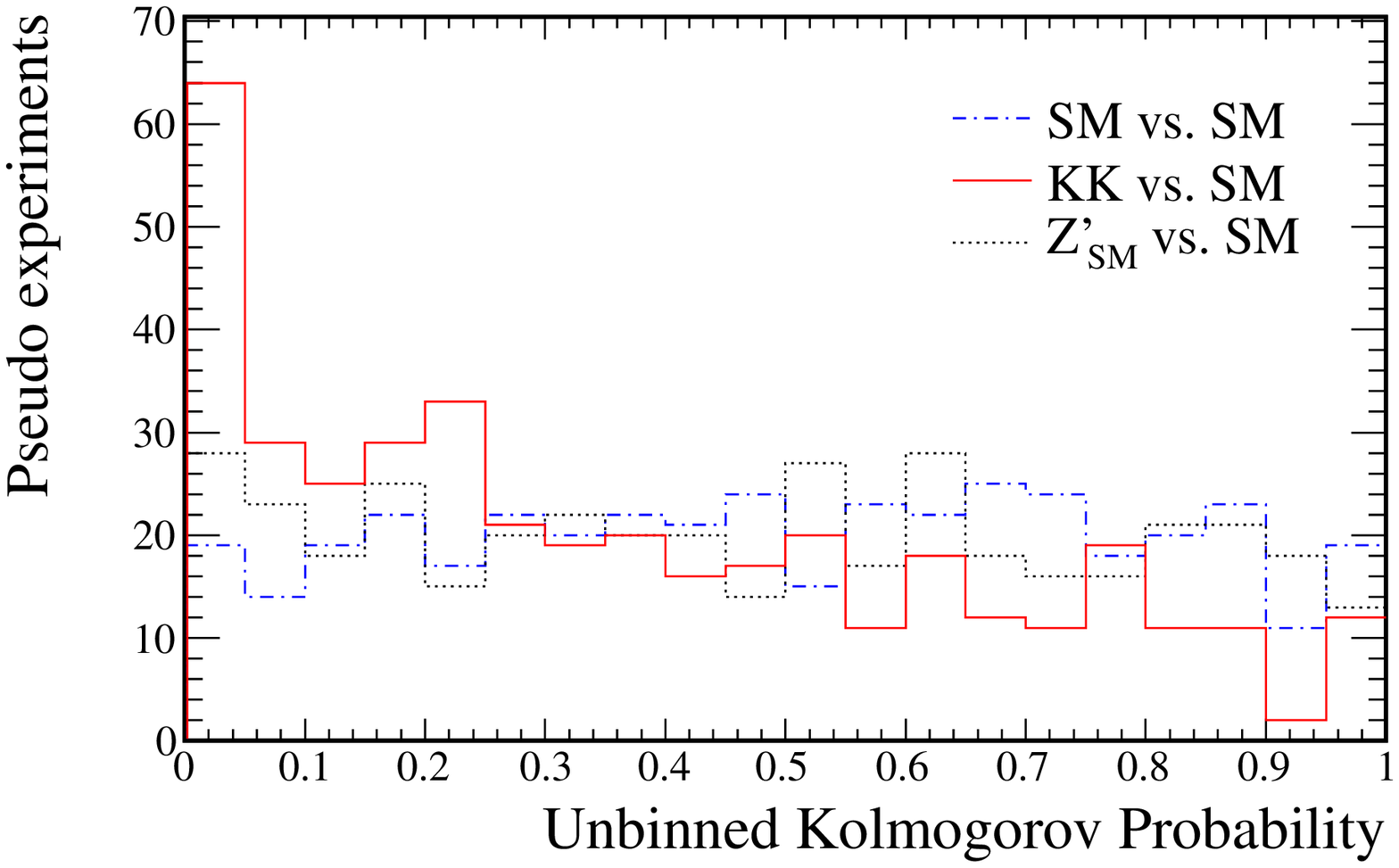}
          \put(60,110){(b) \ $m^*=4$~TeV}\put(60,95){Invariant mass}
        \end{overpic}
      \end{minipage}
      \begin{minipage}{7.5cm}
        \begin{overpic}[scale=0.4]{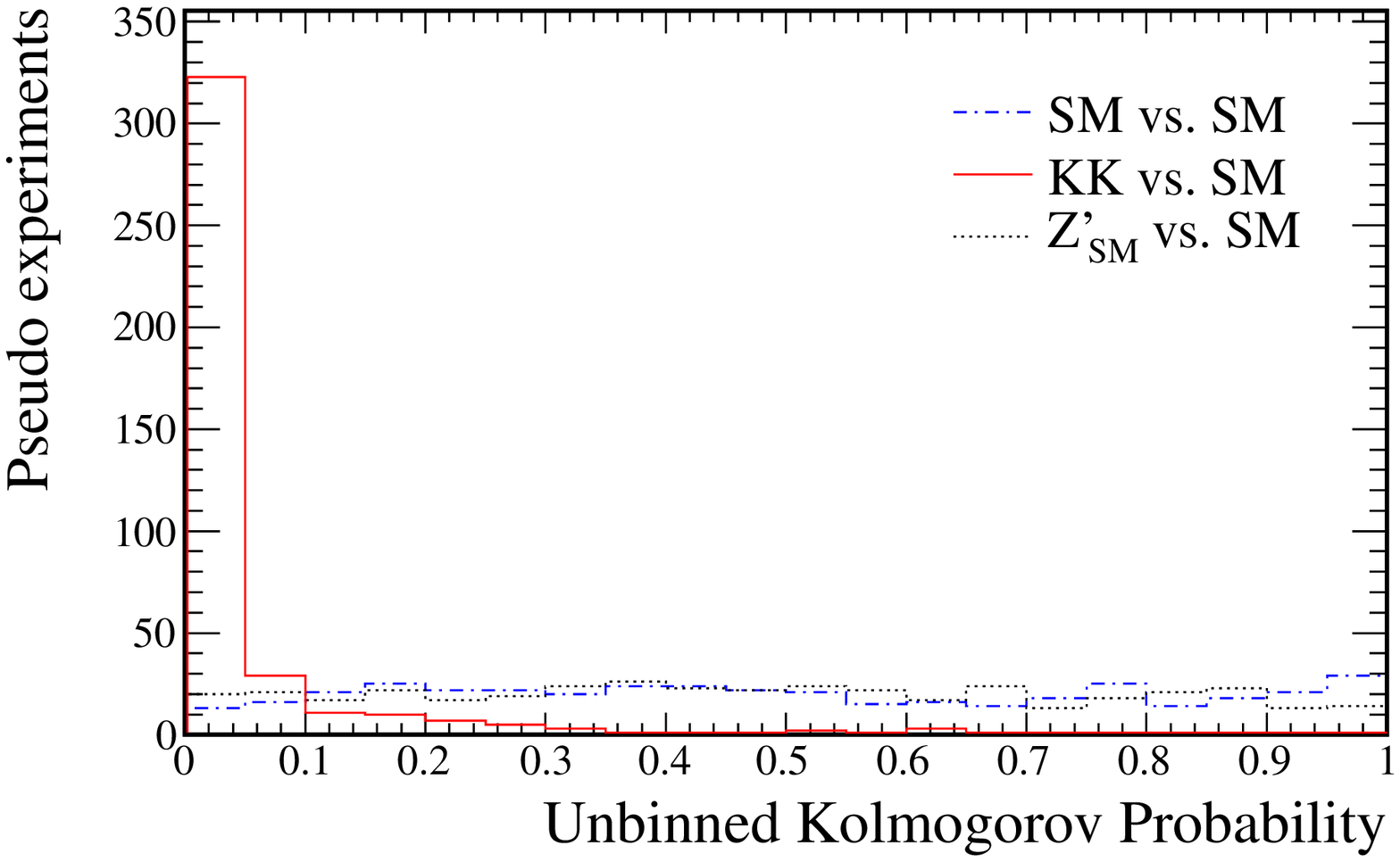}
          \put(60,110){(c) \ $m^*=3$~TeV}\put(60,82){Transverse momentum}
        \end{overpic}
      \end{minipage}
      \hspace{1mm}
      \begin{minipage}{7.5cm}
        \begin{overpic}[scale=0.4]{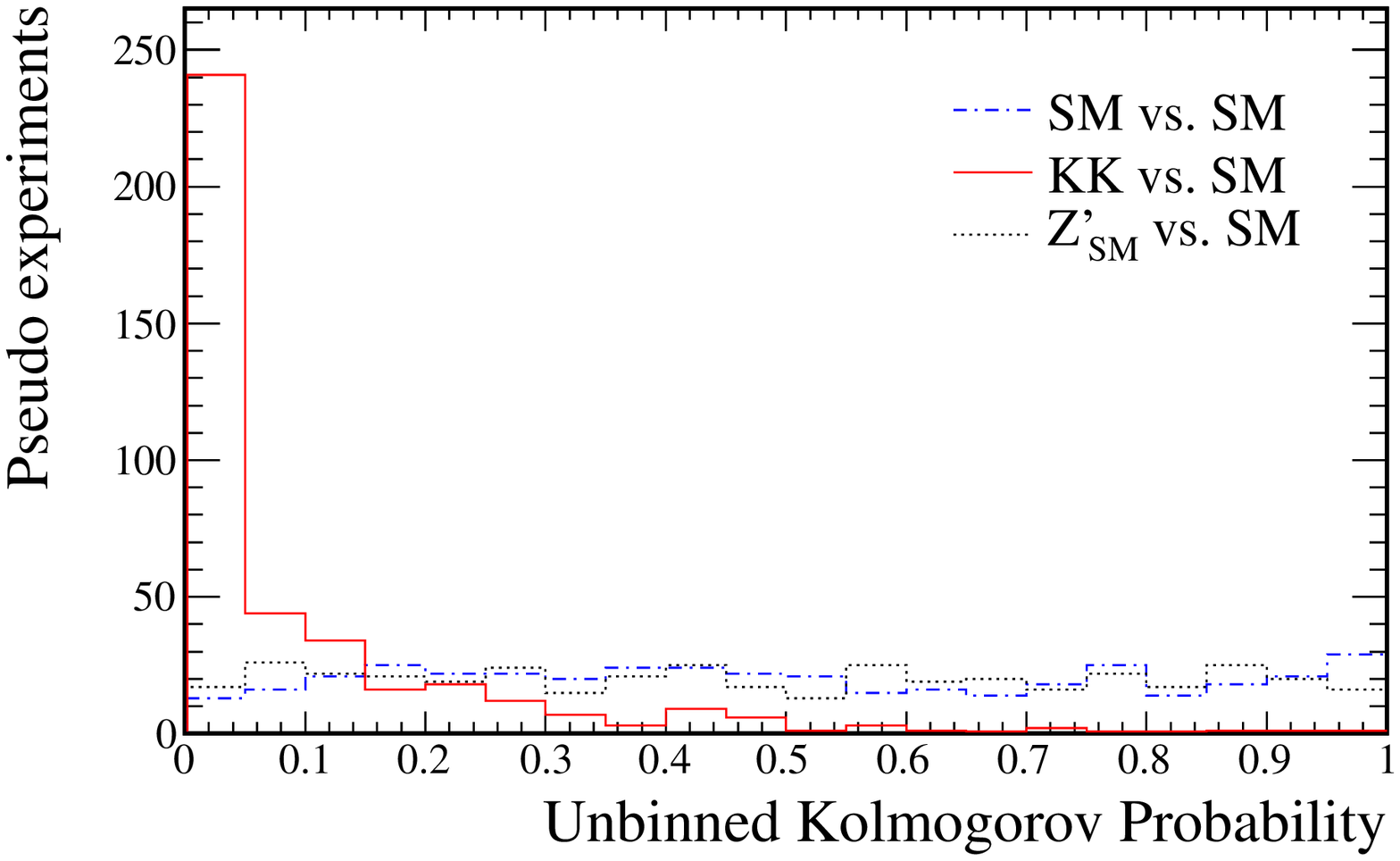}
          \put(60,110){(d) \ $m^*=4$~TeV}\put(60,82){Transverse momentum}
        \end{overpic}
      \end{minipage}
    \end{minipage}
    \caption{
      The distribution of the probabilities of the Kolmogorov statistic for 
      comparison between the data from the 400 pseudo-experiments with that from a large 
      reference sample of SM events for the three models, SM (blue, dot-dashed); KK (solid, red); 
      $Z'_{\rm SM}$ (black, dashed). (a) The distribution for the comparison of the invariant mass 
      distribution with $m^*=3$~TeV, (b) the invariant mass distribution with  $m^*=4$~TeV, (c) and (d) 
      the distributions for the comparison of the muon transverse momentum distribution 
      with $m^*=3$~TeV and $m^*=4$~TeV respectively.
    }
    \label{fig:SMkolmogorov}
  }
}

\newcommand{\figten}{
  \FIGURE[tth]{
    \hspace{-0.6cm}
    \begin{minipage}{15cm}
      \begin{minipage}{7.5cm}
        \begin{overpic}[scale=0.4]{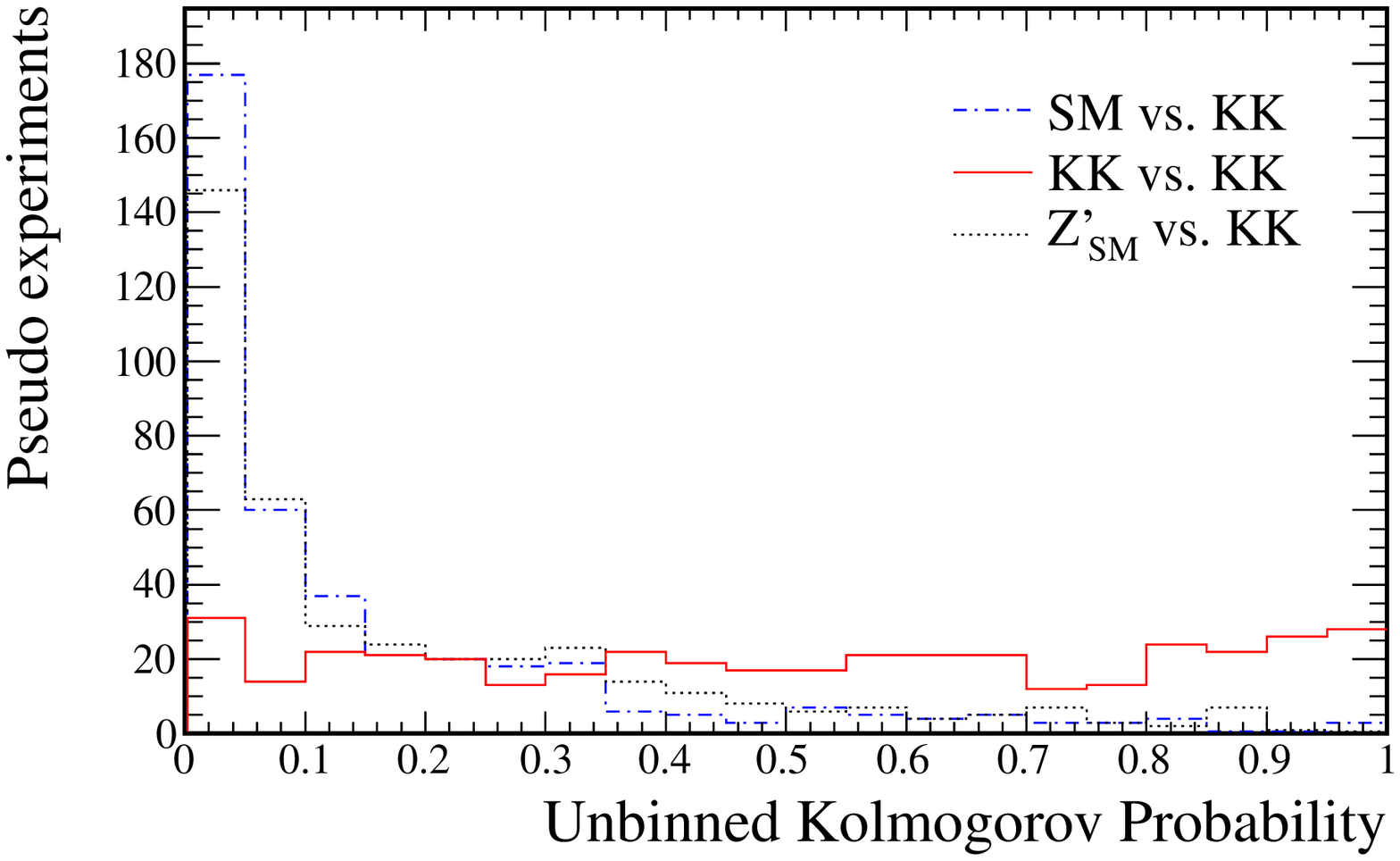}
          \put(60,110){(a) \ $m^*=3$~TeV}\put(60,95){Invariant mass}
        \end{overpic}
      \end{minipage}
      \hspace{1mm}
      \begin{minipage}{7.5cm}
        \begin{overpic}[scale=0.4]{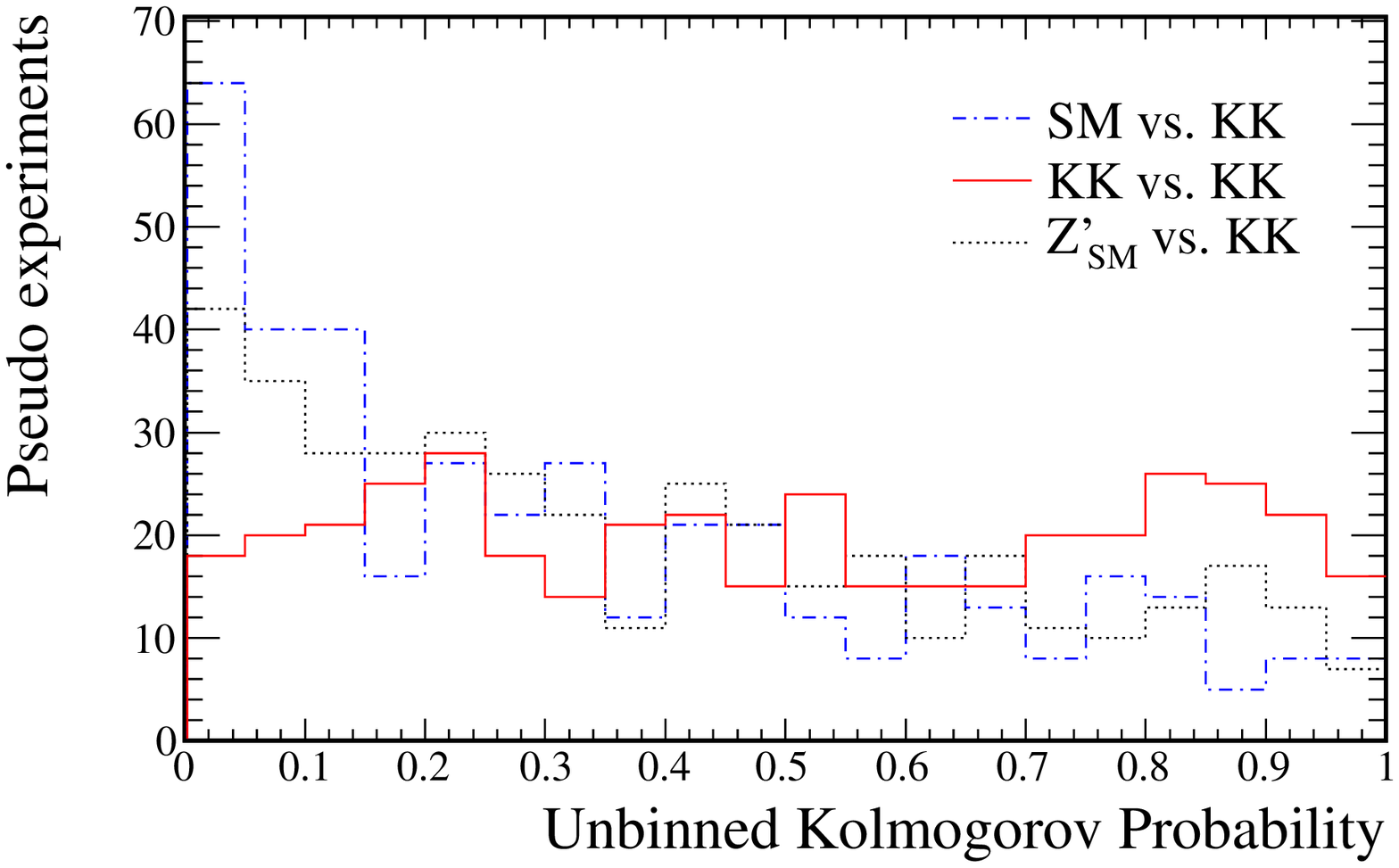}
          \put(60,110){(b) \ $m^*=4$~TeV}\put(60,95){Invariant mass}
        \end{overpic}
      \end{minipage}
      \begin{minipage}{7.5cm}
        \begin{overpic}[scale=0.4]{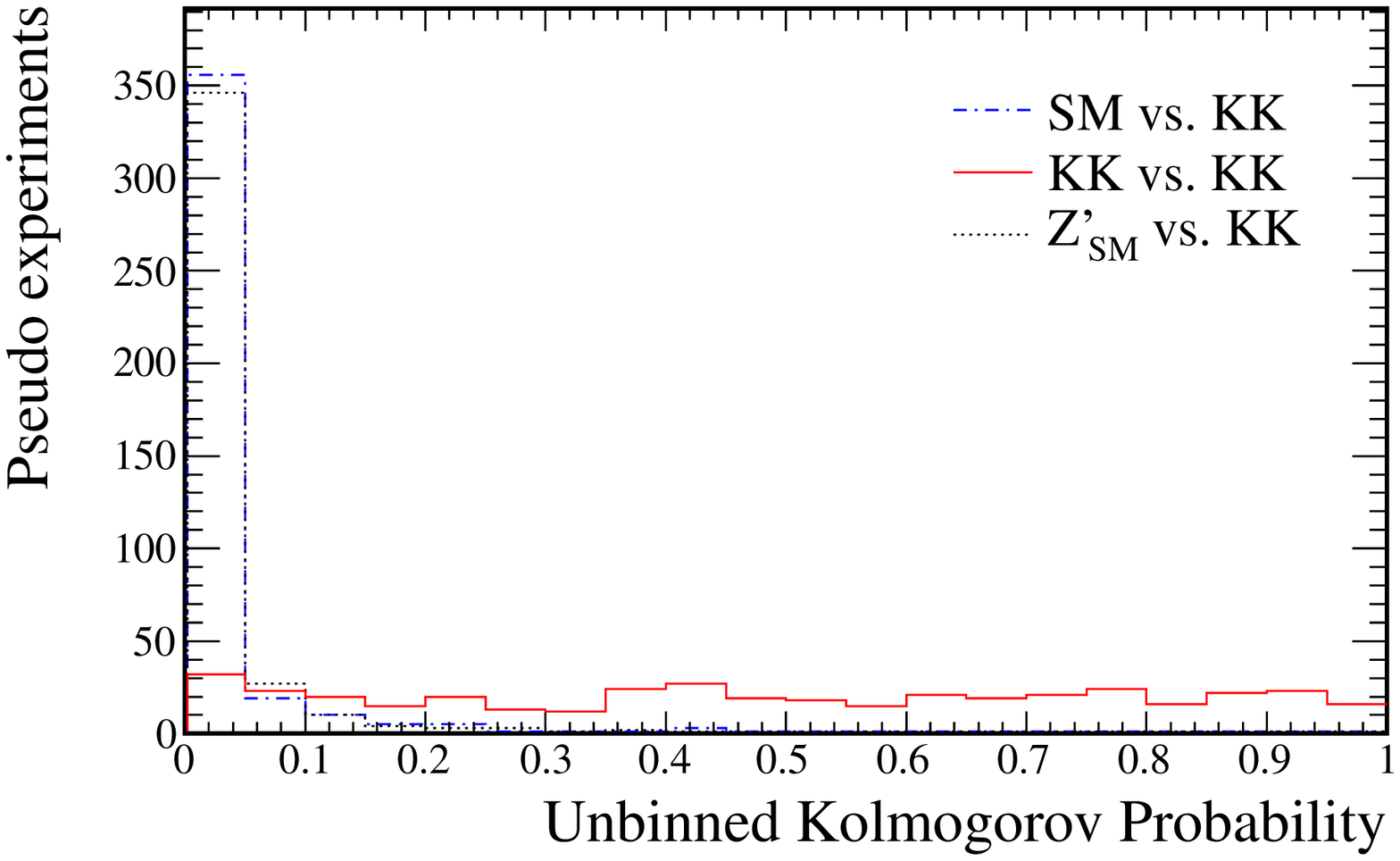}
          \put(60,110){(c) \ $m^*=3$~TeV}\put(60,82){Transverse momentum}
        \end{overpic}
      \end{minipage}
      \hspace{1mm}
      \begin{minipage}{7.5cm}
        \begin{overpic}[scale=0.4]{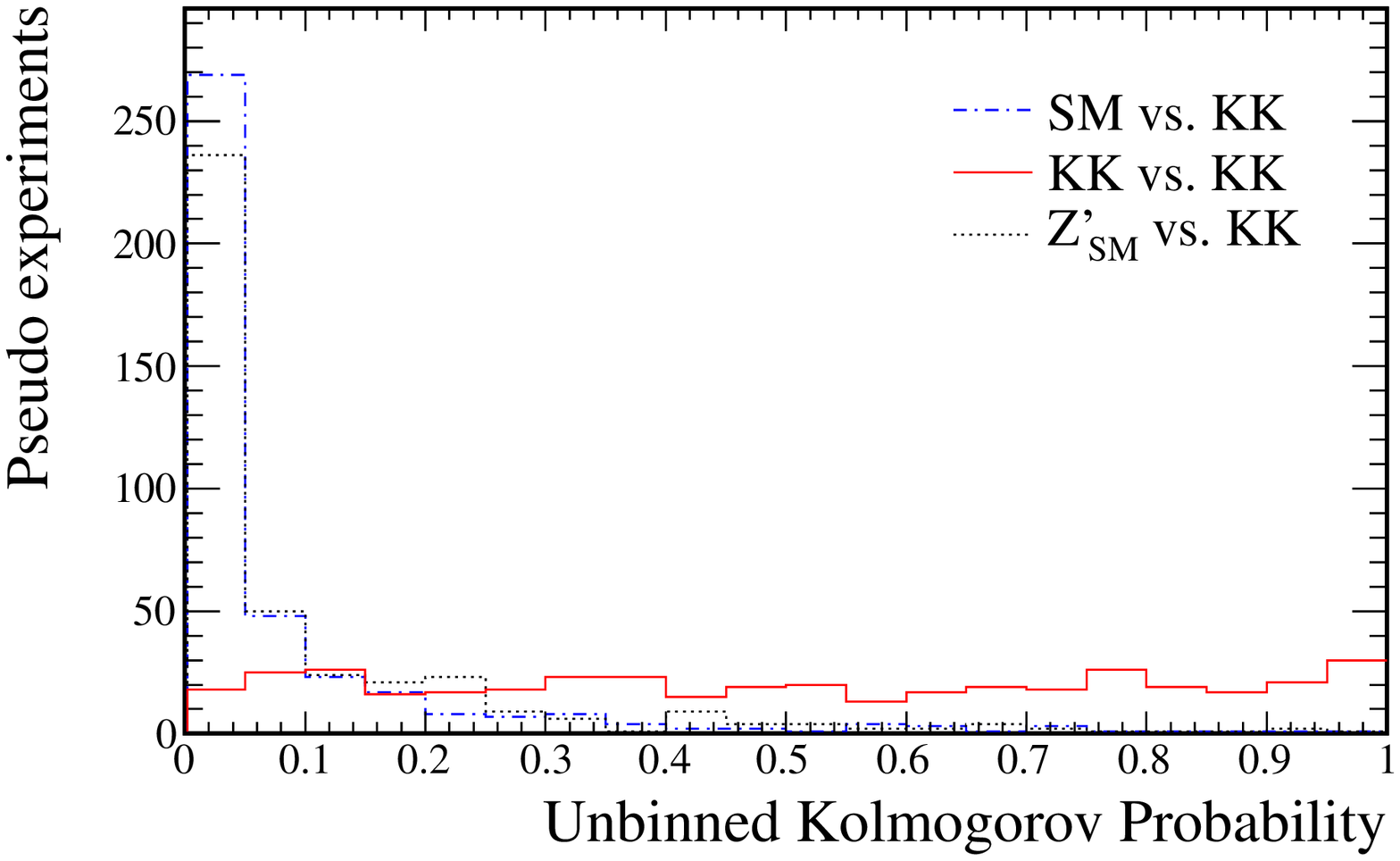}
          \put(60,110){(d) \ $m^*=4$~TeV}\put(60,82){Transverse momentum}
        \end{overpic}
      \end{minipage}
    \end{minipage}
    \caption{
      The distribution of the probabilities of the Kolmogorov statistic for 
      comparison between the data from the 400 pseudo-experiments with that from a large 
      reference sample of KK events for the three models, SM (blue, dot-dashed); KK (solid, red); 
      $Z'_{\rm SM}$ (black, dashed). (a) The distribution for the comparison of the invariant mass 
      distribution with $m^*=3$~TeV, (b) the invariant mass distribution with  $m^*=4$~TeV, (c) and (d) 
      the distributions for the comparison of the muon transverse momentum distribution 
      with $m^*=3$~TeV and $m^*=4$~TeV respectively.
    }
    \label{fig:KKkolmogorov}
  }
}

\newcommand{\tabone}{
\TABLE{%\begin{table}[hhh]
  \begin{tabular}{cccc}
    Model          & $m^*$  &  $\sqrt{\hat s}$ distributions & $p_T$ distributions\\
    \hline
    $Z'_{{\rm{SM}}}$ & 3~TeV  & 0.075   & 0.05~~\\
    $Z'_{{\rm{SM}}}$ & 4~TeV  &  0.07~  & 0.043\\
    KK             & 3~TeV  &  0.36~  & 0.81~~\\
    KK             & 4~TeV  &  0.16~  & 0.6~~~\\
  \end{tabular}  
  \caption{
    Probabilities for excluding, within a 95\% confidence level
    the SM given the observed distributions from the 
    KK and $Z'_{\rm SM}$ samples.
  }
  \label{tabone}
}
}

\newcommand{\tabtwo}{
\TABLE{%\begin{table}[hhh]
  \begin{tabular}{cccc}
    Model          & $m^*$ &  $\sqrt{\hat s}$ distributions & $p_T$ distributions\\
    \hline
    $Z'_{{\rm{SM}}}$ & 3~TeV       & 0.37 & 0.87\\
    $Z'_{{\rm{SM}}}$  & 4~TeV       & 0.11 & 0.59\\
    SM              & (KK, 3~TeV)  & 0.44  & 0.89\\
    SM              & (KK, 4~TeV)  & 0.16  & 0.67\\
  \end{tabular}
  \caption{Probabilities for excluding, within a 95\% confidence level, the KK given the observed distributions from the 
    KK and $Z'_{\rm SM}$ samples.
  }
 \label{tabtwo}
}
}

\begin{document}

% \section{The search for Heavy gauge bosons at the LHC}
\section{Introduction}
Several theories~\cite{ARKANIHAMED,ANTONIADIS,RS,DDG} predict the existence of extra dimensions (ED) 
in addition to the usual three spatial and one time dimension. These models allow various 
particles to propagate into the extra-dimensional bulk. In this paper a 
model~\cite{DDG,TEV} where a single 
extra spatial dimension is compactified onto an $S^1/Z_2$ orbifold 
is considered, where the radius of the $S^1$-shaped extra dimension is 
denoted by $R$. 

This model allows the KK modes of $SU(2)\times U(1)$ gauge fields to propagate into the 
extra-dimensional bulk, while restricting all the matter fermions to be localized in the 
usual three space dimensions~\cite{ARKANIHAMED,ANTONIADIS,ARKANIHAMED2}.
The process discussed in this paper is of particular interest in the context of physics
at the LHC~\cite{lhc}, 
since the signal, unlike other possible heavy $Z$-like signals,  such as the $Z'$, %(see next paragraphs), 
manifests a strong destructive interference between the KK and the Standard Model (SM) 
states at invariant-masses much lower than the mass of the first KK resonance itself.
Even for a resonance around $\sim$4~TeV, this suppression is expected to be within the 
reach of the LHC and it could be observed relatively early during LHC 
operation~\cite{ZPRIMEPHENOMENOLOGY,RIZZOMATRIX,RIZZOINTERFERENCE}.

\newcommand{\noam}{
Using either \mosesWithSpace and \pythia\ or \pythiaWithSpace alone to generate 
the new KK Monte Carlo (MC) events, all the various steps of the event generation, not 
relating to the hard process itself, are done by \pythiaNoSpace.
This includes all parts of the complete event, {\it e.g.}, Initial (Final) State 
Radiation (ISR (FSR)) effects, generation of partons from the incoming protons, 
parton showering, hadronization, proton remnant fragmentation, particle decay etc.
}

For this paper, the analytic form~\cite{RIZZOMATRIX}
for the cross section of the $f\bar f \to \sum_n\left(\gamma^*/Z^*\right)_n \to F \bar F$ 
hard process has been independently verified at leading order by the authors. 
For the subsequent studies, fully simulated events at the hadron level have been 
produced using the \mosesWithSpace framework~\cite{MOSESARXIVE} for the generation of 
the hard subprocess which has been implemented as an external 
process within the \pythia\ Monte Carlo generator~\cite{PYTHIA8}. All the stages in the 
event simulation not involved in the hard subprocess generation, {\em ie}
initial and final state 
radiation (ISR and FSR), generation of the partons kinematics for the incoming 
beam protons, parton showering, hadronization, proton remnant fragmentation
and particle decay etc. are performed by \pythia. %\ framework. 

In a previous study~\cite{KKEXPERIMENTAL} for this KK process the
matrix elements were interfaced to the {\sc Fortran} {\sc Pythia}-v6 
and the generated events passed through the fast simulation~\cite{atlfast}
of the {\sc ATLAS} detector~\cite{atlas}, assuming proton-proton 
interactions at \slhc=14~TeV centre-of-mass energy, integrated LHC luminosity, 
\llhc=100~fb$^{-1}$, and a KK resonance mass of $m_{Z^*}=4$~TeV. % {\em NEED TO THINK ABOUT WHAT TO SAY FOR THE REST OF THIS PARAGRAPH.}
% The results discussed here are consistent with the ones presented 
% at~\cite{KKEXPERIMENTAL} where here, we look at the early LHC program of 
% operation and the derived possibilities. 
Previous studies of the possible $Z'$ signal at the LHC have also been performed~\cite{ZPRIMESTUDIESINLHC}.
%~\cite{ZPRIMETOE+E-,ZPRIMESTUDIESINLHC}.

This paper replicates aspects of the previous study with independently verified 
Matrix Elements and also considers the additional model which assumes the 
existence of an extra heavy, $Z'$, boson arising from a spontaneous breaking 
of a higher gauge symmetry group~\cite{ZPRIMEPHENOMENOLOGY,ZPRIMEPHYSICS}.
%With this additional 
In this additional model, $Z'$ bosons can be produced with different couplings. %in different scenarios.
One scenario often introduced is the $Z'_{\rm{SM}}$ where the new boson has 
the same couplings as the $Z^0$ but with different mass and width. 
% One should note that there 
While there is no theoretical preference for the choice of SM-like couplings for the $Z'$,
distinguishing between the KK and $Z^\prime$ cases where the couplings 
are SM-like is % a study done using several other $Z'$ possibilities, shows that this is 
experimentally more challenging. % (see section 3).

In addition, a new quantitative analysis of several Monte Carlo pseudo-data sets, generated using 
\mosesWithSpace and \pythia, is presented in order
to study the discovery potential for the observation of heavy Kaluza-Klein 
gauge bosons as a consequence of recent expected early LHC running scenarios. % In particular, 
In contrast to the previous study~\cite{KKEXPERIMENTAL} which concentrated 
on masses around a resonance at 4~TeV, with 14~TeV centre-of-mass energy and a large integrated 
luminosity of 100~fb$^{-1}$, this analysis also 
considers the effect of the lower 
7~TeV centre-of-mass energy and  the reduced assumed integrated luminosity of 10~fb$^{-1}$
to ascertain what can be achieved using 
the lower invariant mass region below the resonance,  accesible with the early running scenarios.
%
% In order to enhance the potential of this data with lower centre-of-mass energy instead 
% the full 
% of looking only around the KK resonance as in the previous study, 
% the focus here is on much lower center of mass energy region that is expected to be 
% achievable during the early running scenarios.
%
The effects of initial and final state radiation from 
the incoming partons or outgoing leptons are also discussed.

% and the prospects for the observation of a KK signal with the lower beam 
% energy and luminosity of the early LHC program are considered. 

% Where this study reproduces aspects of the earlier study~\cite{KKEXPERIMENTAL}
% it is found to be in agreement.

\section{The general Kaluza-Klein hard process at leading order}
For this study the process 
$q \bar{q} \to \sum_n\left(\gamma^*/Z^*\right)_n \to l \bar{l}$ was implemented in {\tt C++}
where $q$ and $\bar{q}$ are incoming quark and anti quark, and $l$ and $\bar{l}$ are outgoing leptons, 
however, for generality, the following discussion is presented in terms of $f \bar f$ and $F \bar F$ 
where $f(F)$ can be any SM initial (final) state fermion. 
% For the LHC, the index $f$ should be replaced by $q$ for the initial state $q$ or $\bar{q}$ partons. 
At tree-level, the formulation for the differential cross section for the process 
$f \bar f  \to \sum\limits_{n}\left(\gamma^*/Z^*\right)_{n} \to F \bar F$ can be written as
\begin{equation}
%	\frac{d\hat \sigma \left(\hat s,\cos \theta^* \right)}{d\hat t} = \frac{2}{\hat s} 2\pi \frac{\alpha_{\rm em}^2}{4\hat s}\frac{N_C^F}{N_C^f}\frac{\hat s^2}{4}\sum\limits_{\lambda_f = \pm \frac{1}{2}} {\sum\limits_{\lambda_F = \pm \frac{1}{2}} {\left| \sum\limits_{n=0}^\infty M_{\lambda_f \lambda_F}^{\left(n\right)} \right|^2 \left( {1 + 4\lambda_f \lambda_F \cos \theta^* } \right)^2 } }
\frac{d\hat \sigma \left(\hat s\right)}{d\cos \theta^*} = 2\pi \frac{\alpha_{\rm em}^2}{4\hat s}\frac{N_C^F}{N_C^f}\frac{\hat s^2}{4}\sum\limits_{\lambda_f = \pm \frac{1}{2}} {\sum\limits_{\lambda_F = \pm \frac{1}{2}} {\left| \sum\limits_{n=0}^\infty M_{\lambda_f \lambda_F}^{\left(n\right)} \right|^2 \left( {1 + 4\lambda_f \lambda_F \cos \theta^* } \right)^2 } }
\label{eq:sigmatotalamplitude_KK}
\end{equation}
where $\hat{s}$ is the squared invariant mass of the $q\bar{q}$ state, $N_C^{f(F)}$ is 
the number of colors of $f(F)$, $\lambda_{f(F)}$ is the helicity of the $f(F)$ fermion
and $\cos\theta^*$ is the cosine of the scattering angle with respect to the incoming fermion 
direction  of the outgoing fermions in the $f\bar{f}$ rest frame.
The complete amplitude consists of an infinite tower of Kaluza-Klein excitations 
with increasing mass,
\begin{equation}
\sum\limits_{n=0}^\infty M_{\lambda_f \lambda_F}^{\left(n\right)} \equiv M_{\lambda_f \lambda_F} + \sum\limits_{n=1}^\infty M_{\lambda_f \lambda_F}^{\left(n\right)},
\label{eq:onlytotalamplitude_KK}
\end{equation}
where the SM term ($n=0$) is,
\begin{equation}
M_{\lambda_f \lambda_F} \equiv \frac{e_f e_F}{\hat s} + \frac{{g_{\lambda_f } g_{\lambda_F} }}{{\hat s - m_{Z^0}^2 + i \hat s \frac{\Gamma_{Z^0}}{m_{Z^0}}}},
	\label{eq:totalamplitude}
\end{equation}
and the contribution of the $n^{\rm th}$ excitation for $n=1,2,3,...$ can be written as,
\begin{equation}
	M_{\lambda_f \lambda_F}^{\left(n>0\right)}\left(\hat s \right) \equiv 
	\frac{e_f^{(n)} e_F^{(n)}}{\hat s - {\left(m_{\gamma^*}^{\left(n\right)}\right)}^2 + i \hat s \frac{\Gamma_{\gamma^*}^{\left(n\right)}}{m_{\gamma^*}^{\left(n\right)}}} +
	\frac{g_{\lambda_f}^{\left(n\right)} g_{\lambda_F}^{\left(n\right)}}{\hat s - {\left(m_{Z^*}^{\left(n\right)}\right)}^2 + i \hat s \frac{\Gamma_{Z^*}^{\left(n\right)}}{m_{Z^*}^{\left(n\right)}}}.
	\label{eq:totalamplitude_KK}
\end{equation}
\noindent
The SM helicity couplings~\cite{ZLINESHAPE} of the $Z^0$ to the incoming and outgoing 
fermions are,
% \begin{equation}
%		g_{\lambda_f} =
%			\left\{
%				\begin{array}{rl}
%					-\frac{ e_f \sin^2 \theta_W}{\sin \theta_W \cos \theta_W} &\mbox{for $\lambda_f=+1/2$}\\\\
%					\frac{I_f^3 - e_f \sin^2 \theta_W}{\sin \theta_W \cos \theta_W} &\mbox{for $\lambda_f=-1/2$}
%				\end{array} \right.
%	\label{eq:couplings}
%\end{equation}
\begin{equation}
\label{eq:couplings}
g_{\lambda_f}=   %\cases{
\left\{
  %\begin{cases}
  \begin{array}{ll}
   %-\frac{ e_f \sin^2 \theta_W}{\sin\theta_W\cos\theta_W} & \text{\rm for $\lambda_f=\frac{1}{2}$}\\
  \vspace{2mm}
    \displaystyle{-\frac{e_f\sin^2\theta_W}{\sin\theta_W\cos\theta_W}} & \mbox{\rm for $\lambda_f=+\displaystyle{\frac{1}{2}}$}\\
    \displaystyle{\frac{I^3_f-e_f\sin^2\theta_W}{\sin\theta_W\cos\theta_W}}  & \mbox{\rm for $\lambda_f=-\displaystyle{\frac{1}{2}}$}\\
   \end{array}
  %\end{cases}
\right.
\end{equation}
where the couplings of the KK states to fermions are larger than their SM counterparts 
(equation~\ref{eq:couplings}) by a factor of $\sqrt{2}$~\cite{KKEXPERIMENTAL,RIZZOPEDAGOGICAL}.
The $n^{\rm th}$ KK excitation masses $m_{Z^*}^{(n)}$ and $m_{\gamma^*}^{(n)}$ are given by,
\begin{equation}
\begin{array}{rl}
	&m_{Z^*}^{(n)} = \sqrt{m_{Z^0}^2 + (n\cdot m^*)^2}\\
	&m_{\gamma^*}^{(n)} = n \cdot m^*.
\end{array}
\label{eq:KKmasses}
\end{equation}
where the KK mass, $m^*$, is dependent on the extra dimension
size, $R$, through the relation $m^* \equiv R^{-1}$. The current, indirectly obtained 
theoretical lower bound for $m^*$ assuming that there are no other beyond-the-SM 
(BSM) effects besides the KK model, is 
% around 4~TeV~\cite{KKEXPERIMENTAL,KKCONSTRAINTS,KKEWGLOBALFIT,PDG}.
around 4~TeV~\cite{KKEXPERIMENTAL,KKCONSTRAINTS,PDG}.
% \noindent
The total decay width of the KK $Z^*$ appearing in Eq~\ref{eq:totalamplitude_KK}, 
is given by,
\begin{equation}
	\Gamma_{Z^*}^{(n)} = \Gamma_{Z^0} \times 2\frac{m_{Z^*}^{(n)}}{m_{Z^0}} + \Gamma_{Z^* \to t\bar t}^{(n)},
\label{eq:KKZwidth}
\end{equation}
where $\Gamma_{Z^* \to t\bar t}^{(n)}$ is calculated separately due to the mass of the top quark,
\begin{eqnarray}
  \Gamma_{Z^* \to t\bar t}^{(n)} & = & 
  2 \frac{ N_C^t G_{\mu} m_{Z^0}^2 m_{Z^*}^{(n)} } {24\pi\sqrt{2}} 
  \left[1-\frac{4m_t^2}{\left(m_{Z^*}^{(n)}\right)^2}\right]^{\frac{1}{2}}\\
  & \ & \ \ \times \ \ \left[ 
    1
    -\frac{4m_t^2}{\left(m_{Z^*}^{(n)}\right)^2} 
    + \left(2I_t^3 -4e_t \sin^2\theta_W\right)^2 
    \left(1+\frac{2m_t^2}{\left(m_{Z^*}^{(n)}\right)^2}\right)  \right]
\label{eq:Z2ttbarwidth}
\end{eqnarray}
The total decay width of the (massive) KK $\gamma^*$ appearing in Eq~\ref{eq:totalamplitude_KK}, is,
\begin{equation}
	\Gamma_{\gamma^*}^{(n)} = \sum\limits_{F \neq t}\frac{N_C^F \alpha_{\rm em} m_{\gamma^*}^{(n)}}{6}  \times
		\left\{
			\begin{array}{rl}
				0         &\mbox{for $n=0$} \\
				4e_F^2    &\mbox{otherwise}
			\end{array} \right\} + \Gamma_{\gamma^* \to t\bar t}^{(n)},
\label{eq:KKgammaWidth}
\end{equation}
where the sum is over all the fermionic decay channels, $F\bar F$ except for $t\bar t$, assuming SM channels 
only and where $\Gamma_{\gamma^* \to t\bar t}^{(n)}$ is,
\vspace{-0.2cm}
\begin{equation}
	\Gamma_{\gamma^* \to t\bar t}^{(n)} = 2\frac{ \alpha_{\rm em} N_C^t m_{\gamma^*}^{(n)} }{6} 2e_t^2 
        \left[1-\frac{4m_t^2}{\left(m_{\gamma^*}^{(n)}\right)^2}\right]^\frac{1}{2}
	\left[1 + \frac{2m_t^2}{\left(m_{\gamma^*}^{(n)}\right)^2} \right]
\label{eq:gamma2ttbarwidth}
\end{equation}
Equations~\ref{eq:sigmatotalamplitude_KK} and~\ref{eq:onlytotalamplitude_KK} represent 
a large tower of interfering contributions at increasing masses
%\footnote{In both \mosesWithSpace and \pythiaNoSpace, this is handled by using 
% the C++ standard \shellfont{complex} class, with all the benefits of the familiar complex operations.}.
% The described KK tower is demonstrated in figure~\ref{fig:KK_feynman}.
and is illustrated in figure~\ref{fig:KK_feynman}.

\figone
For the remainder of this paper, only the di-muon final state is considered so the 
index $F$ shall be replaced by $\mu^-$.  
% The next-to-leading order calculation for the KK process is not currently available. 

\section{The KK signal at the LHC}
The KK processes considered here, 
% $\sum\limits_{n}\left(\gamma^*/Z^*\right)_{n}\} \to \mu^+\mu^-$
$pp \to \{\gamma/Z^0$, $Z'_{\rm SM}\ { \rm or}\ \sum_n\left(\gamma^*/Z^*\right)_n\} \to \mu^+\mu^-$
exhibits several outstanding characteristics that can provide a strong suggestion 
of the presence of the first KK resonance even before it is directly 
seen~\cite{RIZZOINTERFERENCE}.
These characteristics also enable discrimination between other similar 
possible signals, such as the  $Z'_{{\rm{SM}}}$, assuming that the signal
can be directly observed~\cite{MOSESARXIVE,EVGENYTHESIS}.
In this paper an attempt is made to address the applicability of such statements % based on unique KK signatures, 
in the light of the likely running scenario for the updated LHC schedule
in terms of the available collider centre-of-mass energy and the 
luminosity. % ~\cite{citation}. 

The next-to-leading order (NLO) corrections to $Z'$ production have recently been 
calculated~\cite{QUACKENBUSH} and have been shown to be around 25\% of the 
leading order contribution. Since the vector-axial vector couplings of the 
standard model, KK and $Z'$ bosons are all proportional, it might be expected
that the NLO correction to the KK signal is also of the same 
magnitude and an NLO K-factor of 1.25 to scale the leading order 
cross section to that predicted by the full NLO  prediction
could be applied. Since applying such an 
overall factor after generation of events, will not affect the statistical 
precision of the predictions 
such a correction has not been applied to any Monte Carlo predictions of the 
cross section, but it should be noted that any observed signal in the data, 
might be expected to be 25\% higher than the Monte Carlo predictions presented
here. 

\figtwo

The principal difference between all the $Z'$ models and the KK model is the lack 
of a heavy photon in the $Z'$ models. The missing interference terms of such a heavy
photon in the Standard Model and $Z^\prime$ amplitudes modify the 
\shat, $p_T(\mu^-)$ and $\cos\theta^*$ distributions. 
Although the $Z'_{{\rm{SM}}}$ choice is not theoretically preferred, 
it is in practice the most challenging for comparison. A study~\cite{RIZOZPRIME}
%~\cite{anothercitation} 
of the relevant $Z'$ scenarios suggests that while they behave similarly at invariant 
masses below the resonance, those resonances 
with the non SM couplings are in general narrower 
and smaller than the $Z'_{\rm SM}$
which is already narrower and smaller than the KK resonance. 
Moreover, as with the KK resonance, the $Z'_{\rm SM}$ also introduces a suppression 
of the cross section at invariant masses below the resonance - 
although this suppression is small - whereas the 
other $Z'$ scenarios do not~\cite{RIZOZPRIME}. Therefore, since it is the 
closest in shape to the rather wide KK resonance, the case where the couplings 
of the $Z'$ are SM-like, will be the most difficult to distinguish from the KK case.

% , and for the remainder of this 
% paper it should be assumed all cross sections might be 25\% higher 
% than shown. For any detailed statistical comparison, an increase in statistical 
% precision would make any differences more apparenent so any values 
% quoted should be taken as a conservative estimate.
%%  although for consistency with the KK signal, 
%%  available at leading order only, these have not been included in the 
%%  subsequent analysis. 
%% Concurrently, the $Z'_{\rm SM}$ option is the easiest to distinguish from the 
%% SM.

For the remainder of this section, an overview of these characteristics 
is given for $\sqrt{s}_{{\rm{LHC}}}=14$~TeV and for a KK $m^*$ or $Z^\prime_{\rm SM}$ mass of 4~TeV. 
All the results will be given within the acceptance for the typical general purpose 
LHC detector and trigger for muons, 
{\it ie}, $p_T>$10~GeV, $\left|\eta\right|<$2.5.
For illustrative purposes, large Monte Carlo reference samples were generated using \mosesWithSpace
and \pythiaWithSpace to the level of the full, final state hadrons, 
for each of the SM, $Z'_{\rm SM}$, and KK models.
Samples %of pseudo-data
corresponding to an LHC integrated luminosity of 
100~fb$^{-1}$ were also produced.
A discussion in the context of the lower LHC centre-of-mass energy, and lower 
integrated luminosity of the initial LHC running programme with 
various values of the KK mass parameter, $m^*$, is given 
in section~\ref{sensitivity}.

At least three unique signal characteristics of the KK process  with respect to 
either the SM or the $Z'_{{\rm{SM}}}$ model can be derived from the di-muon invariant 
mass distribution (line-shape), the muon $p_T$ and $\cos \theta^*$ distributions. 

The invariant mass distributions of the three models can be seen in 
figure~\ref{fig:theory_3models}(a) for the nominal signal masses of 4~TeV mentioned earlier.  
The strong suppression of the cross section  for the KK 
line-shape with respect to the SM is clearly seen for masses below half the mass 
of the first KK resonance. Note that the $Z'_{{\rm{SM}}}$ does not differ from the SM 
line-shape as strongly as the KK line-shape and that this difference is generally 
to increase the cross section with respect to the SM expectation.

\figthree

From the $\cos \theta^*$ distribution  a clear difference can be 
seen between the KK and the $Z'_{\rm SM}$ forward-backward asymmetries, 
$A_{fb}$, defined as
\begin{equation}
A_{fb} \equiv \frac{N_f-N_b}{N_f+N_b}
\label{AFB}
\end{equation}
where $N_f$ and $N_b$ are the number of events before the acceptance cuts on 
the outgoing leptons, for which respectively, $\cos\theta^* > 0$ (forward) and 
$\cos\theta^* < 0$ (backward). 
%% 
%% Since it is not known which beam hadron contains the incoming quark and which the anti-quark
%% in practice, $\cos \theta^*$ cannot be measured with respect to the incoming quark direction.
%% In addition, the symmetry of the LHC beams, requires that 
%%

Due to the symmetry of the LHC beams, it is not known which beam proton contains the 
incoming quark and which the anti-quark, so $\cos\theta^*$ cannot be measured with respect 
to the incoming quark direction. Hence 
for a meaningful definition of the 
asymmetry, some  event-by-event definition of direction with respect to which the 
angle of the outgoing muon is measured must be adopted. 
At  leading order  % {\em (SHOULD WE QUOTE A FRACTION??)} 
the di-muon boost direction will in general coincide with that of the incoming quark along 
the $z$ direction since the valence quarks are, on average, more energetic than the anti-quarks 
which originate entirely from the sea. 
As such, the asymmetry, $A_{fb}^{\beta}$, is defined, measured with respect 
to the di-muon boost direction, $\vec{\beta}$, in the lab frame. 
Under this definition, $\cos\theta^*_{\beta}$ is now the cosine of the angle between the 
outgoing muon and the boost direction.
% and the corresponding asymmetry, denoted by $A_{fb}^{\beta}$ was chosen. % , as defined in~\cite{MOSESARXIVE}. 

% The effect in the asymmetry of the between the KK and 
% the $Z'_{\rm SM}$ very little sensitivity to the replacement of $\theta^*$ 
% with $\theta '$ in the asymmetry calculation.

The theoretical asymmetry, $A_{fb}^{\beta}$, as a function of the di-muon invariant mass
of the three processes can be seen in figure~\ref{fig:theory_3models}(b) 
% as a function of the di-muon $\cos\theta^*$ with respect to the boost direction in the lab frame 
and shows a clear difference  between the KK and the $Z'_{{\rm{SM}}}$ asymmetries around the 
4~TeV mass of the simulated signal states. 
%
% At Next-to-Leading Order (NLO),

\figfour

Radiation from the initial state partons 
or the final state leptons will in general lead 
to the correspondence between the boost direction of the di-muon system and the  
quark direction being lost to some degree since the di-muon state itself 
can obtain some transverse momentum with respect to the beam line. 
As such, it may be better to measure $\cos\theta^*$ in the Collins-Soper 
frame~\cite{COLLINSSOPER} where the $\hat{z}$ axis is defined along the 
bisector of the beam directions in the di-lepton rest frame, 
with the positive direction 
taken as that closer to $\vec{\beta}$. In this frame, the angle of
the muon % $\mu^-$
with respect to the positive $\hat{z}$ axis is denoted by 
$\theta'$.  
% and in this frame is denoted by $\cos\theta^\prime$.

The differential distributions for the large reference Monte Carlo samples of the processes implemented using 
\mosesWithSpace and \pythia\ can be seen in figure~\ref{fig:KK_kinematics_highLuminosity}
where events were generated using the MRST parton distribution set~\cite{MRST}.   
The invariant mass distribution for each of the three models is shown in 
figure~\ref{fig:KK_kinematics_highLuminosity}(a), the muon transverse momentum and 
pseudorapidity distributions are shown in 
figures~\ref{fig:KK_kinematics_highLuminosity}(b) and \ref{fig:KK_kinematics_highLuminosity}(c)
and finally the distribution of  $\cos\theta'$
% the cosine of the centre-of-mass scattering angle of the di-muon 
% system in the Collins-Soper frame with respect to the boost direction in the lab frame 
is shown in figure~\ref{fig:KK_kinematics_highLuminosity}(d). To enhance the forward-backward 
asymmetry, for  the events in figure~\ref{fig:KK_kinematics_highLuminosity}(d), an addition 
requirement of $2\leq\shat\leq 5~$TeV has been applied. 

In the distribution with respect to $p_T$, a significant enhancement can be seen above 
the SM expectation for transverse momenta around 2~TeV for both the KK and $Z'_{{\rm{SM}}}$ 
models. However, the $p_T$ enhancement for the KK process 
begins approximately 500~GeV before that from the  $Z'_{\rm SM}$ and it is significantly 
larger. 

In the kinematic region around the potential signal states, the large invariant mass of the 
lepton pair resulting from the decay of the heavy state, will in general result in each lepton possessing 
a very large transverse momentum. Since the  mass reconstruction of the intermediate state is 
dependent on the muon reconstruction it will in general be less precisely measured for muons 
with high transverse momentum.

\figfive

In order to study the effect on the observation of any potential signal resulting from the finite 
detector resolution the outgoing muon momentum has been smeared by 
\[
\sigma(p_T) = 0.12 \cdot p_T {\rm[TeV]}
\]
typical of the resolution obtainable by either the ATLAS of CMS experiments~\cite{ATLASRES,CMSRES} for $\sim$1~TeV muon tracks.
The reconstructed invariant and transverse momentum distributions for ``reconstructed'' detector level 
muons can be seen in figure~\ref{fig:threeb}, after the same "true" tracks used to obtain the distributions seen in figure~\ref{fig:KK_kinematics_highLuminosity}, had been smeared.
In this case, clearly the invariant mass peaks and the enhancement 
in the transverse momentum distribution have been greatly smeared, notably in the very high 
invariant masses or transverse momenta, although the characteristic valley in the invariant mass
distribution is clearly seen.

In \pythia, the effects of initial and final state radiation from the incoming 
partons or outgoing muons can be easily controlled by switches. Their influence on 
the di-muon system, can be seen by considering the transverse momentum, $Q_T$, of the 
di-muon system. 
% of the $\gamma^*/Z^*$ or the $Z'_{{\rm{SM}}}$ {\em **** NB: We need to check this, shouldn't 
% $Q_T$ be tbe transverse momentum of the dimuon system? If it is the transverse momentum 
% of the intermediate state then FSR should not affect $Q_T$, only ISR. **** }
Figure~\ref{fig:QT} shows the the distribution of %the di-muon transverse momentum, 
$Q_T$ versus the di-muon invariant mass, \shat\ for the case where
initial- and final-state radiation is switched off, figure~\ref{fig:QT}(a) and where
they are both switched on,  figure~\ref{fig:QT}(b).
This clearly shows that for the case where an initial parton has radiated the di-muon 
system will in general be boosted in the transverse plane to balance the radiation
for any given di-muon invariant mass. 
% and so the di-muon state itself will be boosted. 
For the case of radiation from one of the final state leptons, the intermediate state 
will not in general have significant transverse momentum but the di-muon state itself will be 
balanced by the transverse momentum of the radiated photon which may not be observed
or associated with the di-muon system.

Both of these effects will therefore have an impact on the $\cos\theta^*_{\beta}$  
and $p_T$ distributions. Radiation of a final state photon 
which is not reconstructed as part of the di-muon system 
will give rise to a reduction of the measured di-muon invariant 
mass, \shat, which will not be present 
for the case of initial state radiation.

\figsix

Figure~\ref{fig:KK_kinematics_lowLuminosity100} shows an example of the achievable statistics, 
possible with the design LHC yearly integrated luminosity of 100~fb$^{-1}$ at 14~TeV, for fully 
simulated events at the generator level, including the effects of initial and final state radiation
and the effects of the smearing of the muon momentum to simulate the finite detector 
resolution.  
In particular, it can be seen that the general features of the KK signal discussed earlier 
remain visible under these conditions. Moreover, it is expected that the deviations from the 
Standard Model expectation in the \shat\ and $p_T(\mu)$ 
distributions will be sufficient 
to suggest the existence of a KK resonance even if the mass of the resonance is beyond the reach of the 
LHC.

% {\em ----------  SUTT EDITED DOWN TO HERE -------------}

\section{The LHC sensitivity for the KK signal}
\label{sensitivity}
The design luminosity and centre-of-mass energy of the LHC are not expected to be 
achieved before the 2012-2013 shutdown.
%~\cite{SomeStatement}. 
This imposes strong limitations on the discovery and 
identification potential at the TeV scale for the first few years of LHC operation. 
For the early LHC running period, 
the distributions for the variables seen in figure~\ref{fig:KK_kinematics_lowLuminosity100} 
for \slhc=14~TeV and \llhc=100~fb$^{-1}$ will be significantly different. Even 
during early LHC operation however, it might be possible to observe, at lower masses, the 
remote shadow that a much heavier KK state casts in the foothills of the resonance.
The  current bound on $m^*$ is around 4~TeV, obtained by indirect 
% methods~\cite{IndeirectBounds}, 
methods~\cite{PDG}, 
but for completeness, masses lower than this have also been considered for this study.

\figseven

To study the discovery potential for early LHC running scenarios, the na\"{i}ve significance for 
observing the KK peak and the characteristic dip at invariant masses below the resonance has 
been studied. The LHC centre-of-mass energies considered were 
\slhc=7~TeV and 
\slhc=14~TeV with integrated luminosities in the  range
\llhc=1~fb$^{-1}$ to \llhc=100~fb$^{-1}$. 
For each LHC configuration the significance, $S(m^*)$, for observing 
either the peak or the valley, defined by 
\begin{equation}
% 	S\left(m^*\right) = \frac{\left|N_{\rm BSM}\left(m^*\right) - N_{\rm SM}\right|}{\sqrt{N_{\rm SM}}}
	S(m^*) = \frac{\left|N_{\rm BSM}\left(m^*\right) - N_{\rm SM}\right|}{\sqrt{N_{\rm SM}}},
\label{eq:significance}
\end{equation}
was calculated for various values of $m^*$, 
where $N_{\rm BSM}$ and $N_{\rm SM}$ are the number of expected beyond-the-SM  and SM events, 
and where BSM stands for either KK or $Z^\prime_{\rm SM}$. For this calculation, an overall 
K-factor of 1.25 has been applied to both the BSM and SM cross section.

\figeight

The unique form of the KK line shape enables the strict definition of integration ranges 
for different values of $m^*$. This feature is present throughout the entire KK tower, 
where the $n^{\rm th}$ KK peak will always be at $\sqrt{\hat s} \simeq n m^*$ between 
$m^*\left(n - \frac{1}{2}\right)$ and $m^*\left(n + \frac{1}{2}\right)$.
As such, the expected number of events for a given integrated luminosity can be  calculated by 
integrating over the invariant mass distributions in either of two mass ranges;
\begin{itemize}
\item the KK peak region;  $\frac{1}{2}m^* \leq \shat \leq \frac{3}{2}m^*$, {\em ie.} 
between the first two adjacent KK local minima on either side of the first KK mode,
\item the KK valley region;  $2m_{Z^0} \lesssim \shat \leq \frac{1}{2}m^*$, {\em ie.} 
from approximately $\sim$200~GeV up to the first KK local minimum.
\end{itemize}

In figure~\ref{fig:significance}, the significance in these peak and valley regions is shown 
as a function of the BSM mass. For illustration the $5\sigma$ limit is also shown.
Within the models studied here, it can be seen that the significance for the Kaluza-Klein model
is always larger than that of the $Z'_{\rm SM}$ model, as would be expected from the smaller
signal.
% \footnote{Again,  is this just an artifact of the $Z'$ parameters we have chosen? Could they be larger??}

\fignine

There are two cases where the significance is above $5\sigma$ level and is 
higher for the KK valley region than for the KK peak region. These are 
\begin{enumerate}
\item \slhc=7~TeV, \llhc=10~fb$^{-1}$ and for $2.5 \lesssim m^* \lesssim 3$~TeV.
\item \slhc=14~TeV, \llhc=100~fb$^{-1}$ and for $m^* \gtrsim 5$~TeV,
\end{enumerate}
whereas for all other cases, the significance of the KK peak region is larger. 
This indicates that even when the mass of the KK state is beyond the kinematic 
limit and cannot be observed directly, the suppression of the cross section at 
lower invariant masses may still provide useful information.

However, in the integration over the invariant mass distribution, without 
a sensitive choice of the limits of the integration which are not {\em a priori} known, 
there will be some cancellation between the number of events from the peak
and the valley regions which may reduce the apparent significance.
% However, this picture is only partially correct since by calculating the number of events, 
% $N_{BSM}$ or $N_{SM}$, part of the information is lost.

A more sensitive method in this case may be to use the  shape of the invariant mass distribution 
itself rather than obtain $N_{\rm BSM}$ and $N_{\rm SM}$ by integrating over some unknown peak and valley 
regions.
%which are in any case not known {\em a priori}.
% In this case, it is expected that comparing the line shapes will 
% enable to identify the strong KK suppression with respect to the SM expectation.
% The last statement is most relevant for the first LHC years of operation, in the case where 
This might be most useful for the data from the first few years of LHC 
operation where both the beam energy and luminosity will be lower. In such 
a case, although a resonance itself might not be observed, the first 
signs of a faster than expected fall of the cross section may provide an indication 
of a signal in the TeV region.

\figten

To ascertain the likely potential for discrimination between the various models 
using the early LHC data, data from 400 pseudo-experiments was generated with three 
datasets per pseudo-experiment - one each for the Standard Model, the KK and $Z'_{\rm SM}$ 
models, including the muon transverse momentum smearing to simulate the effects of finite
detector resolution. An analysis comparing the results of each pseudo-experiment to a large 
reference sample was performed using the reconstructed $\mu^+\mu^-$ invariant mass and the 
observed muon transverse momentum distributions.
Figure~\ref{fig:binbybin} illustrates the distributions for the data from a single 
pseudo-experiement for the case of \slhc=7~TeV and \llhc=10~fb$^{-1}$ and two 
possible values of $m^*=3$~TeV and $m^*=4$~TeV. 
% Figures~\ref{fig:binbybin}(a) and figure~\ref{fig:binbybin}(b)
% show the invariant mass distributions for the $m^*=3$~TeV and 4~TeV cases respectively, and 
% figures~\ref{fig:binbybin}(c) and figure~\ref{fig:binbybin}(c) show the same but for the muon 
% transverse momentum distribution. 

In figure~\ref{fig:SMkolmogorov} the results of an unbinned Kolmogorov probability comparison
of % the each of 
the three data sets in each psuedo experiment with a large SM reference sample is 
presented for the invariant 
mass distribution 
and for the
transverse momentum distribution of the leading muon.
Each comparison has been performed for pseudo-data samples 
with \mbox{$m^*=3$~TeV} % (Fig.~\ref{fig:SMkolmogorov}a) 
and \mbox{$m^*=4$~TeV}. %(Fig.~\ref{fig:SMkolmogorov}b)
   For the distributions used for these  comparisons, 
   the leading order cross sections have been assumed. 
%   Applying an overall K-factor of 1.25 after event generation, to provide an appoximation
%   of the full NLO order cross sections
%   would increase the cross sections, but not the statistical precision 
%   or descrimination between the samples and so such a correction has not been applied.
   Since the NLO K-factor of 1.25 would predict a larger cross section than the LO 
   cross section presented, this might suggest a small increase in both 
   the statistical precision and discrimination between the various signals would be 
   possible with actual data, and as such the
   results presented here should be considered as a
   conservative estimate.

%
%also for $m^*=4$~TeV 
%(Fig.~\ref{fig:SMkolmogorov}c) 
%and $m^*=3$~TeV. %(Fig.~\ref{fig:SMkolmogorov}d). 
% Each figure 
% shows the probability for 
% obtaining that value for the Kolmogorv statistic for each of the pseudo-data sets.
%
%
%
In all four distributions from figure~\ref{fig:SMkolmogorov} the probablility 
for obtaining the Kolmogorov statistic  
when comparing  the small SM pseudo-datsets with the large 
SM sample is flat, as expected for compatible data sets, as is that for the $Z'_{\rm SM}$ model. 
% in both the invariant mass and transverse momentum distributions at 
% both boson mass scales. 
The comparison of the KK signal is however, 
quite heavily peaked at lower probabilities, being most significant 
for the transverse momentum distribution, most notably for the $m^*=3$~TeV
scale. 
%, suggesting a significant, observable difference between the KK and SM signals,
%even with these low numbers of events. 

Similarly, figure~\ref{fig:KKkolmogorov} shows the same comparison of the 
three data sets for each pseudo-experiement, but this time compared to a large sample
of KK events. In this case, the distributions of probabilities from comparing the KK samples
are flat, again as expected for compatible distributions, whereas both the SM and $Z'_{\rm SM}$ 
samples are peaked at lower probabilities.

\tabone

This indicates a potentially observable incompatibility between the KK 
pseudo-\linebreak[4]experiments and the large SM reference sample. The probabilities 
to exclude the SM hypothesis with a 95\%~confidence level 
for the KK or $Z'_{{\rm{SM}}}$ samples  presented here can 
be seen in table~\ref{tabone}. The probabilities to exclude the KK hypothesis with a 
95\%~confidence level given the SM or $Z'_{{\rm{SM}}}$ models are presented in 
table~\ref{tabtwo}. 

\tabtwo

This suggests that even with the reduced luminosity and centre-of-mass energy available 
during the early LHC running, it may still be possible to distinguish a low mass KK signal 
from both the Standard Model, and a $Z'_{\rm SM}$ signal similar to that studied. 

% \tableone

\section{Summary and Outlook}
The hard process for simulating the tower of Kaluza-Klein electroweak boson exchange within 
the $S^1/Z_2$ extra dimensional model has been independently verified, and implemented 
in the \mosesWithSpace framework and integrated with the \pythiaWithSpace %(v8.140) 
generator.
The process itself is also now included~\cite{torbjorn} in the most recent release of \pythia.
This process is particularly interesting 
in the context of the early LHC programme of operation since it features several signatures that 
might enable the observation of a significant departure from the corresponding Standard Model 
expectation.
A preliminary study considering the observation of
evidence for a resonance inside or just beyond the kinematic reach 
of the LHC has been performed for the early LHC running scenario, and suggests 
that for a Kaluza-Klein resonance above 3~TeV such an observation may still be 
possible with early LHC data, whereas for a model based on the $Z'_{\rm SM}$ 
such an observation in the same mass range may not be possible.

%% A preliminary study of the possibilities for distinguishing between the Kaluza-Klein  and 
%% $Z'_{\rm SM}$ signals, assuming an observed resonance with masses greater than 1~TeV has been 
%% performed for the early LHC running scenario and shows that for lower mass KK states below 
%% 2.5~TeV discrimation 
%% between the Kaluza-Klein model and Standard Model may still be possible and discrimination 
%% of a signal at 4~TeV with a 14~TeV collision centre-of-mass energy may be possible with only 
%% 1~fb$^{-1}$.

% done in~\cite{MOSESARXIVE}.
% The results obtained there, show that this is possible only at $\sqrt{s}_{{\rm{LHC}}}=14$~TeV 
% and above $\mathcal{L}_{{\rm{LHC}}}=100$~fb$^{-1}$. To quantify the sensitivity for different 
% LHC run conditions and assuming that no resonance is observed, a corresponding analysis should 
% be done with fully simulated events.  

% \acknowledgements{Acknowledgments}
\section{Acknowledgments}

The authors would like to  acknowledge the financial support of the Israeli Science Foundation, the United Kingdom 
Science and Technology Facilities Council and the MCnet 
programme.
Personal thanks are given to % the \href{http://www.montecarlonet.org/}{MCnet} people and mainly to 
Mike Seymour and Jonathan Butterworth for supporting the work on the {\sc Moses} framework. % \mosesWithSpace 
% framework~\cite{MOSESARXIVE}. 
Special thanks are given to the {\sc Pythia} authors,  
Torbj\"{o}rn Sj\"{o}strand and Stefan Ask for useful discussions and 
advice on the integration of the discussed KK hard process within the 
\pythia\ framework.
The authors would also like to thank Evgeny Yurkovsky and Tom Rizzo for fruitful discussions.

%Our thanks are given to Evgeny Yurkovsky\footnote{The Raymond and Beverly Sackler School of Physics \& Astronomy, Tel Aviv University} for sharing the results from his study on this model.
%In addition, our thanks are given to \href{http://www.isf.org.il/english/default.asp}{ISF} and to \href{http://www.montecarlonet.org/}{MCnet}.
%Special thanks are given to Torbj\"{o}rn Sj\"{o}strand, Mike Seymour and Jonathan Butterworth.

%\pagebreak[4]
% \section*{References}

\end{document}